\documentclass[prl,nofootinbib,preprint,superscriptaddress, 11pt]{revtex4}
\usepackage[usenames,dvipsnames]{color}
\usepackage{graphicx}
\usepackage{amsmath}
\usepackage{amsfonts}
\usepackage{amssymb}
\usepackage{dsfont}
\usepackage{dcolumn}
\usepackage{bm}
\usepackage{slashed}
\usepackage{pstricks}

\usepackage{slashed}
\usepackage{multirow}
\usepackage{array}
\usepackage{rotating}
\usepackage{xcolor}
\usepackage{epstopdf}
\usepackage{fancybox}

\usepackage{ulem,fancyvrb}

\newcolumntype{x}[1]{
>{\centering}p{#1}}%
\newcommand{\tnhl}{\tabularnewline\hline}

\newcommand{\GeV}      {~\mathrm{GeV}}
\newcommand{\pb}      {~\mathrm{pb}}

\def \cha{\widetilde{\chi}^{\pm}_1}
\def \chb{\widetilde{\chi}^{\pm}_2}
\newcommand{\beqn}{\begin{eqnarray}}
\newcommand{\eeqn}{\end{eqnarray}}
\newcommand{\be}{\begin{equation}}
\newcommand{\ee}{\end{equation}}
\newcommand{\non}{\nonumber \\}

\newcommand{\mathsym}[1]{{}}

\def \cha{\tilde{\chi}^{\pm}_1}
\def \chb{\tilde{\chi}^{\pm}_2}

\def \na{\tilde{\chi}^{0}_1}
\def \nb{\tilde{\chi}^{0}_2}
\def \nc{\tilde{\chi}^{0}_3}
\def \nd{\tilde{\chi}^{0}_4}
\def \n34{\tilde{\chi}^{0}_{3,4}}
\def \g{\tilde{g}}
\def \ql{\tilde{q}_L}

\def \ta{\tilde{t}_1}

\def \ba{\tilde{b}_1}

\def \sta{\tilde{\tau}_1}

\def \smr{\tilde{\mu}_R}
\def \ser{\tilde{e}_R}

\def \snl{\tilde{\nu}_{\tau}}

\def\transph{S_T\geq 0.2}
\def\met100{\slashed{E}_T\geq 100 \GeV}

\def \smr{\tilde{\mu}_R}
\def \ser{\tilde{e}_R}

\def \snl{\tilde{\nu}_{\tau}}

\newcommand{\st}{Stueckelberg~}

\begin{document}

\title{\Large  Low Mass Gluino  within the Sparticle Landscape, Implications for  Dark Matter, and  Early Discovery Prospects at LHC-7}

\author{Ning Chen}
\affiliation{C.N.\ Yang Institute for Theoretical Physics, 
Stony Brook University, Stony Brook, NY 11794, USA}
  
\author{Daniel Feldman}
\affiliation{Michigan Center for Theoretical Physics,
University of Michigan, Ann Arbor, MI 48109, USA}

\author{Zuowei Liu}
\affiliation{C.N.\ Yang Institute for Theoretical Physics, 
Stony Brook University, Stony Brook, NY 11794, USA}

\author{Pran Nath}
\affiliation{Department of Physics, Northeastern University,
 Boston, MA 02115, USA.}

\author{Gregory Peim}
\affiliation{Department of Physics, Northeastern University,
 Boston, MA 02115, USA.}

\preprint{MCTP-10-49; YITP-SB-10-34; NUB-TH-326}


\begin{abstract}
We analyze  supergravity models that predict a low mass gluino within the landscape of
sparticle mass hierarchies. The analysis includes 
a broad class of models  that arise in minimal and in 
nonminimal supergravity unified frameworks and in extended models
with additional  $U(1)^n_X$ hidden sector gauge symmetries.
 Gluino masses in the range $\left(350-700\right)$~GeV are investigated.  Masses in this range are 
promising for early 
discovery at the  LHC  at $\sqrt s =7$~TeV (LHC-7).
The models  exhibit a wide dispersion in the gaugino-Higgsino eigencontent of their LSPs and in their associated sparticle  
mass spectra. A signature analysis is carried out and the prominent discovery channels for the models are identified 
with most models needing only $\sim 1~\rm fb^{-1}$  for  discovery at  LHC-7. 
In addition, significant variations in the discovery capability 
of the low mass gluino models are observed for models in which the gluino masses are 
of comparable size due to the mass splittings in different models  and the relative position of
the light gluino
within the various sparticle mass hierarchies. 
The models  are 
consistent with the current stringent bounds from the Fermi-LAT,
 CDMS-II,  XENON100, and  EDELWEISS-2  experiments.    A subclass of these models, which 
 include a mixed-wino LSP and a Higgsino LSP, are  also shown to accommodate
 the positron excess seen in the PAMELA satellite experiment. 
 \\\\\\\\\\\\\\\\ Published online: February 4, 2011, Phys. Rev. D 83, 035005 
 \end{abstract}

\maketitle

\section{I. Introduction}
The exploration of supersymmetry at colliders within the landscape of
sparticle mass hierarchies can  
provide insight into the nature of the underlying theory\cite{land1,land2,land3,land4,land5,Land8,land6,land7} .  
Thus, in the minimal supersymmetric standard model there are 32 supersymmetric
particles including the Higgs bosons and in general there  exists a mass hierarchy among them.  
One  can estimate the number of  sparticle mass  patterns that may appear by use of 
Stirling's formula ($n!\sim \sqrt{2\pi n} (n/e)^n$), which gives close to $10^{28}$  different possibilities  including
sum rules for the sparticle masses. Limiting oneself to the first four lightest sparticles and assuming that the lightest supersymmetric particle (LSP) is the neutralino, the number  is 
significantly reduced but is still around  $\mathcal{O}(10^{4})$. On the other hand, in supergravity unified
models with radiative  breaking of electroweak symmetry, the number of allowed possibilities is much smaller\cite{land1,land2,land3}. Further, it has been  demonstrated in several subsequent 
works 
 \cite{land2,land3,land4,land5,Land8,land6,land7}, 
 that analyses of supersymmetry from the point of view of  the  possible low mass sparticle  hierarchies indeed shed
light on the underlying theory   as the specific  mass orderings   and mass gaps 
among sparticles  dictate the type  of  signatures  that  would be produced at the Large Hadron Collider. 
Specifically, in this work we investigate the landscape under the constraint that the 
gluino has a low mass and  is discoverable in early runs at the LHC at $\sqrt s =7$~TeV and with a few fb$^{-1}$ of integrated
luminosity (for a recent review  see \cite{Nath:2010zj}).

Because of their strong color interactions, 
gluinos and squarks are expected to be 
produced copiously  in high energy  $pp$ collisions at the LHC,
and  a wide class of supergravity models are capable of such sparticle production.
Several works have already appeared
which point to encouraging results  for the possibility of discovery of supersymmetry 
in the early data  \cite{flnearly,Baer:2010tk,Peim,Feldman:2010uv,Giudice:2010wb,Wacker,moreearly,Chen:2010yi}.
Here, our analysis of sparticle mass hierarchies is done within the context of supergravity 
grand unified 
models\cite{sugra} where 
  we  include both universal  and nonuniversal soft breaking  \cite{sugra,NUSUGRA}, \cite{land1,land2,land3,NUSUGRAColliders}
that allow for the possibility of 
 low mass gluinos in the mass range of approximately $\rm (350-700) GeV$, with sfermions that could be either light or heavy,
 consistent with all current experimental 
constraints.
In addition, we study extended 
supergravity models where the relic density is satisfied by coannihilations with matter in the 
hidden sector  \cite{fln2,FLNN} 
which is another possible particle physics solution, beyond a Breit-Wigner Enhancement\cite{BWE,recentBWE}
and other mechanisms, 
to achieving a large flux of dark matter in the halo consistent with the relic density.
Dark matter annihilations
in a universe with a nonthermal cosmological history is a  possible  solution as well  \cite{MoroiRandall, Feldman:1900zz}. 

The class of models we  
consider within the context of sparticle mass hierarchies are those which are  
promising  for early SUSY discovery.
A number of  models  
 with low mass gluinos and their  hierarchical mass patterns are exhibited,  and their   LHC signatures at $\sqrt s=7$~TeV are
 analyzed and discussed.  
The strong correlations between the hierarchical structure of the sparticle masses \cite{land1} 
and the early discovery prospects at LHC as well as in  dark matter experiments are studied with specific focus on 
the gluino in the mass hierarchy.
It  is shown that most of the models would become visible with $(1-5)~\rm fb^{-1}$ 
of integrated luminosity.
Further, the models are subjected  to the more recent constraints  from the  
Fermi Large Area Telescope
(Fermi-LAT) data  \cite{linesource}
on the monochromatic radiation that can arise  in annihilation of neutralinos into photons,
as well as from the direct detection experiments on the LSP-nucleon spin-independent cross sections  
\cite{CDMS,Aprile:2010um,collaboration:2010ei}(for a recent overview see
\cite{Feng:2010ef})
which are beginning to put more stringent constraints on supersymmetric models.  It is also
shown that a subclass of the models with a  low mass gluino have a neutralino LSP that can satisfy the positron excess as seen in the PAMELA satellite experiment 
\cite{PAMELA}.
The low mass gluino (LG)  models discussed are consistent with the cold dark matter relic density from the WMAP data  \cite{Jarosik2010}, and most lie within the observed WMAP experimental band. 
The sparticle mass hierarchies corresponding to these  LG models are given in Table(\ref{patterns}).

The outline of the rest of the paper is as follows:  In Sec.(II) we discuss
the  experimental constraints on the model classes studied. These constraints 
include the recent limits from    Fermi-LAT
on the  monochromatic radiation  and the upper limit on the spin-independent 
LSP-nucleon elastic scattering cross section.  In Sec.(III) we 
analyze a number of representative model points that 
encompass a wide range of theoretical frameworks which include
minimal supergravity grand unification
(mSUGRA), nonuniversal supergravity  grand unified models (NUSUGRA), and extended supergravity models with 
a  $U(1)^n_X$ hidden sector gauge symmetry. The specific properties of these models,
including their light sparticle spectra,   are  also discussed in this section. A subclass of the models
is shown to be capable of accommodating the PAMELA data. In Sec.(IV)
we discuss the signatures of the models considered here 
at the LHC with $\sqrt s=7$ TeV. Conclusions are given in Sec.(V).

\section{II. Experimental Constraints \label{constraints}}
\label{section_constraints:}
Below we summarize current constraints on supersymmetric models
from collider and dark  matter experiments which we include in the analysis.
 
  {\it 1. Collider Constraints: }
The models we consider are subject to  several accelerator constraints which are sensitive to  new physics.
These include the experimental result 
$\mathcal{B}r(B\rightarrow X_{s}\gamma)=(352\pm23\pm9)\times 10^{-6}$
from the  Heavy Flavor Averaging Group  \cite{Chan:1997bi}
 along with BABAR, Belle, and CLEO,
while for the standard model prediction we use
the next to next to leading order (NNLO) estimate
of $\mathcal{B}r(b\rightarrow s\gamma)=(3.15\pm 0.23) \times 10^{-4}$~\cite{Misiak:2006zs}. 
Supersymmetry makes important corrections to this process  \cite{susybsgamma}, 
and the current (small) discrepancy between
the experiment and the standard  model predictions 
hints at the presence of  low-lying sparticles  \cite{Chen:2009cw}.
For this analysis we take a $3\sigma$ corridor around the experimental value: $2.77\times 10^{-4}<\mathcal{B}r(b\rightarrow s\gamma)<4.27\times 10^{-4}$. 
Another important flavor constraint is the rare decay process $B_{s}\rightarrow\mu^{+}\mu^{-}$.
We take the recent $95\%$ C.L. constraint by CDF~\cite{CDFB}
 $\mathcal{B}r(B_{s}\rightarrow\mu^{+}\mu^{-})<5.8\times 10^{-8}$. In the context of a 
large spin-independent cross section in the minimal supersymmetric standard model (MSSM), which can occur at low Higgs masses, and large $\tan \beta$, this constraint has recently been shown to be very restrictive on the MSSM parameter space under radiative breaking \cite{land2}, including models with low mass dark matter \cite{Feldman} in the 10 GeV range \cite{Bernabei:2010mq,Aalseth:2010vx}. 
 The correction to $g_{\mu}-2$ is an important indication of new physics, and in models of supersymmetry, significant corrections
 are expected with a low-lying sparticle spectrum  \cite{yuan}.  
 The $g_{\mu}-2$ data  \cite{Bennett:2004pv} 
 has been analyzed recently  
 using improved estimates of the  hadronic correction  \cite{Davier:2009zi}. In this analysis we take  
  a conservative bound: $-11.4\times 10^{-10}<\delta(g_{\mu}-2)<9.4\times 10^{-9}$.
  Experimental limits on the sparticle masses are as discussed in  \cite{Chen:2009cw}.

{\it 2. Dark Matter Constraints: }
In addition to the above, there are constraints from dark matter experiments. 
The seven-year WMAP data  \cite{Jarosik2010} gives 
the  relic density of cold dark matter as  $\Omega_{\rm DM}h^{2}=0.1109\pm 0.0056$. 
For our analysis, we take a conservative upper bound on the relic density predictions in the 
MSSM to take into account the theoretical uncertainties and other possible dark 
matter  contributions (see e.g., \cite{recentBWE}, \cite{Cao:2007fy}, \cite{Hur:2007ur}; 
for early work see \cite{Boehm:2003ha}).  However many of the models produce a relic abundance consistent
with the double-sided WMAP bound at a few standard deviations.
Further, new experimental results from probes of  
direct and indirect detection of dark matter have begun to place more stringent bounds
 on dark matter models. 
Thus, recent results from direct detection experiments CDMS-II  \cite{CDMS},  
XENON100  \cite{Aprile:2010um}, and EDELWEISS-2 \cite{collaboration:2010ei}
 suggest an upper bound to the 
spin-independent (SI) LSP-nucleon scattering cross section.
At present, the limit is $\sigma_{\rm SI}\lesssim 5\times 10^{-8}$~pb 
over the range of interest of the models discussed here.
Many of the models discussed here have spin-independent cross sections 
in the range $\lesssim (10^{-9} -10^{-8}) \pb$ and 
are thus directly relevant to   dark matter  direct detection experiments in the near future. 

Further, the Fermi-LAT experiment
has obtained constraints on the production of  $\gamma$-ray lines with energies from $30$~GeV to $200$~GeV. 
The upper limits of the $\gamma$-ray line flux are in the range of 
$(0.6-~4.5)\times~10^{-9}$~cm$^{-2}$s$^{-1}$  \cite{linesource}. 
This gives rise to 
upper bounds on the corresponding dark matter annihilation cross sections,
and as we will discuss later, the data  constrains the eigencontent of the LSP
 in the mass range analyzed.  Finally, the recent PAMELA  
 \cite{PAMELA} 
 data shows an anomalous high positron flux in the range $(10-100)$~GeV.  The possible 
dominant  contribution to the positron excess in  the MSSM with neutralino dark matter can arise
 from  $\tilde{\chi}_{1}^{0}\tilde{\chi}_{1}^{0}\rightarrow W^{+}W^{-},ZZ$  annihilations, depending
 on the eigencontent of the LSP.   Here, a velocity averaged cross section on the order of
$\langle\sigma v\rangle_{WW+ZZ}  \gtrsim  5\times 10^{-25}$ cm$^{3}$s$^{-1}$  will be shown to provide a
 reasonable explanation of the data without invoking large astrophysical boost factors in the positron flux.
  We will consider the implications of such constraints later in the analysis.

\begin{table}[t!]
\centering
 A sample of low mass gluino models within the landscape\\
\vspace{.3cm}
\scriptsize{
\begin{tabular}{|x{1.2cm}|x{1.2cm}|x{1.2cm}|x{1.2cm}|x{1.2cm}|x{1.2cm}|x{1.2cm}|x{1.2cm}|} 
\hline
 Label & $M_{\tilde{g}}$	& $m_0$	& $M_1$	& $M_2$ & $M_3$& $A_0$	& $\tan\beta$\tnhl
\hline
LG1 & 424 & 2000 & 130 & 130 & 130 & -1000 & 8 \tnhl
LG2 & 715 & 60	& 300	& 300 	& 300	& -100	& 8 \tnhl
	LG3& 386 & 1400	& 800	& 528 	& 132	& 3000	& 25\tnhl
 LG4& 378	& 3785	& 836	& 508 	& 98	& -6713	& 20\tnhl
 LG5 &385 & 2223 & 859& 843 & 130 & 4680 & 48\tnhl
LG6&391 &1180 & 860 & 790 & 138 &2692 & 42 \tnhl
 LG7 & 442	& 2919	& 263	& 151 	& 138	& 4206	& 18\tnhl
LG8 & 417 & 1303	& 257	& 152 	& 139	& 1433	& 18\tnhl
LG9 & 696	& 1845	& 327	& 193 	& 249	& 1898	& 13\tnhl
LG10 & 365 & 1500 & 1600 & 1080 & 120 & 2100 & 15 \tnhl
LG11& 433 & 605 & 302 & 176 & 161 & 1781& 22\tnhl
 LG12& 588 & 636 & 419 & 249 & 228 & 1568 & 37\tnhl
 LG13 & 684 & 335 & 391 & 466 & 279 & -1036 & 3\tnhl
LG14 & 618 & 48 &  289  &  310  & 256  &   -407 & 6\tnhl
 LG15& 602 & 61 &   310 & 351 & 249 &  0 &  9\tnhl
LG16 & 343 & 2200 & 450 & 235 & 100 & 300 & 5 \tnhl
 LG17&425	& 3000	& 400	& 207 	& 125	&   0	& 4\tnhl
\end{tabular}
}
\caption{A sample of low mass gluino models where additionally we take $\mu>0$
and $m_{t(pole)} =173.1~\rm GeV.$  The soft breaking parameters are given at the high scale  of  $\sim 2\times 10^{16}~ \rm GeV$ . 
Nonuniversalites in the gaugino sector $M_{a=1,2,3}$ are taken in 15 of the models. All masses in the
table are in GeV.   A more detailed discussion of the models is given  in the text.  
     } 
 \label{bench}
\end{table}

\begin{table}[h!]
\centering
 A sample of the sparticle landscape for low mass gluino SUGRA models \\
\vspace{0.3cm}
\begin{tabular}{|c||l|c|c|c|} 
\hline
Label	&~~~~~~~~~~~~Mass Pattern & $M_{\g}$ (GeV) & Lightest $M_{\tilde{q}}$ (GeV)  & Gluino Position   \\\hline
LG1 & $\na<\cha<\nb<\g<\nc<\nd$ & 424 & 1985 &  4  \tnhl
LG2 & $\na<\sta<\tilde{\ell}_{R}<\snl<\tilde{\nu}_{\ell}<\tilde{\ell}_{L}$ & 715 & 635 & 31 \tnhl
LG3 & $\na<\g<\cha<\nb<\ta<\nc$ & 386 & 1411 & 2  \tnhl
LG4 & $\na<\g<\cha<\nb<\ta<\nc$ & 378 & 3751 & 2  \tnhl
LG5 & $\na<\g<\cha<\nb<\ta<\sta$ & 385 & 2217 & 2   \tnhl
LG6 & $\na<\ta<\g<\cha<\nb<\nc$ & 391 & 1202 & 3  \tnhl
LG7 & $\na<\cha<\nb<\g<\nc<\nd$ & 442 & 2888 & 4   \tnhl
LG8 & $\na<\cha<\nb<\g<\nc<\nd$ & 417 & 1314 & 4  \tnhl
LG9 & $\na<\cha<\nb<\nc<\nd<\chb$ & 696 & 1882 & 7  \tnhl
LG10 & $\na<\cha<\nb<\g<\nc<\ta$ & 365 & 1511 & 4   \tnhl
LG11 & $\na<\cha\lesssim\nb<\ta<\g<\ba$ & 433 & 690 & 5   \tnhl
LG12 & $\na<\cha<\nb<\sta<\nc<\nd$ & 588 & 790 & 14 \tnhl
LG13 & $\na<\ta<\cha<\nb<\sta<\tilde{\ell}_{R}$ & 684 & 677 & 24   \tnhl
LG14 & $\na<\sta<\tilde{\ell}_{R}<\snl<\tilde{\nu}_{\ell}<\tilde{\ell}_{L}$ & 618 & 550 & 31    \tnhl
LG15 & $\na<\sta<\tilde{\ell}_{R}<\snl<\tilde{\nu}_{\ell}<\cha$ & 602 & 536 & 31 \tnhl
LG16 & $\na<\cha<\nb<\g<\nc<\chb$ & 343 & 2178 & 4  \tnhl
LG17 & $\na<\cha<\nb<\g<\nc<\chb$ & 425 & 2966 & 4  \tnhl
\end{tabular} 
 \caption{
Sparticle mass hierarchies  
 for the low mass  gluino models. Listed are the first six lightest sparticles in the spectra. 
 The lightest squark, shown in column 4, is taken from the first 2 generations.
    In the display of the mass hierarchies we do not include the light CP even Higgs.}
\label{patterns}
\end{table}

\section{III. Models of a Low Mass Gluino and Sparticle Mass Hierarchies \label{benchmarks}}
In $N=1$ supergravity unified models,  gaugino masses can arise from the gauge kinetic  function $f_a$
corresponding to the gauge group $G_a$.
Thus, for the standard model gauge groups $SU(3)_C$, $SU(2)_L$, and $U(1)$ the
gaugino masses are given by
$M_a= (2\mathcal{R}(f_a))^{-1}  F^I \partial_{I} f_a~ (a=1,2,3)$,
where the gauge kinetic function $f_a$ depends on the fields $\phi_I$  and  $F^I$  are
 the order parameters for supersymmetry breaking
  with the VEV of  $F^I$  proportional to the gravitino mass, i.e., $\propto m_{3/2} M_P$.
 The assumption that the
fields $\phi_I$ are singlets of the standard model gauge group will lead to universal gaugino masses.
However, in general the breaking can occur involving both singlet and non-singlet terms
(for a recent analysis see \cite{steve}) and, thus,
the gaugino masses in general will not be universal. Additionally, there can be loop corrections to 
the gaugino masses via exchange of heavy fields where the leading corrections are also
 proportional to $m_{3/2}$, to  the Casimir and to the square of the gauge coupling for
each gauge group (for early work see\cite{loop}). 

A case of immediate interest for LHC analyses is  $M_3(M_U) \in (100-300) \rm GeV$
as the gluino mass at the unified scale ($M_U$) increases in magnitude as one moves towards the electroweak scale  via the renormalization group equations, and  this translates to gluino masses in the range of $\rm (350-750)~GeV$ in the 
model classes considered. While the models considered in this analysis can be obtained with specific
choices of supersymmetry breaking, involving, for example, a combination of singlet and non-singlet F-term
breaking for gaugino masses at the scale $M_U$, we will keep our analysis rather generic in that 
the information on symmetry breaking is contained in the gaugino masses
at the GUT scale which can arise from various models of the gauge kinetic energy 
function, and in general they depend on both the hidden and the visible sector fields. 

Indeed there are several classes of well motivated SUGRA models which lead to sparticle spectra with a low mass gluino.
These include the mSUGRA models  and  nonuniversal SUGRA models  where the soft breaking in the
gaugino sector at the grand unification scale includes non-singlet contributions
as well as singlet  and  non-singlet contributions (see, e.g.,  \cite{steve,GNLSP}). 
Different high scale models can generate different next to lightest supersymmetric particles
(NLSPs) at the electroweak scale, some of which lead to the gluino as the NLSP (GNLSP  models) uncovered in Ref.  \cite{land2}, and further investigated
in \cite{GNLSP,Gogoladze,Wells}. 
We also discuss models where the relic density due to thermal  annihilations is rather small 
but can be 
boosted to the current range of  the WMAP value with coannihilations with matter in the hidden sector\cite{FLNN}.
Such a situation arises in SUGRA models with extended $U(1)^n_X$  gauge groups\cite{FLNN}.  Another 
possibility to boost the relic density arises from nonthermal processes  \cite{MoroiRandall}.  
Both of these model classes, with different types of cosmological histories, 
namely, thermal \cite{FLNN,Chen:2010yi} and nonthermal  \cite{MoroiRandall,Feldman:1900zz}, 
can lead to an  explanation of the positron excess seen by the PAMELA satellite experiment
and provide explanations for the relic abundance of dark matter - however, both classes 
require fields beyond those in the low energy MSSM.

\begin{figure}[t!]
  \begin{center}
 \includegraphics[scale=0.8]{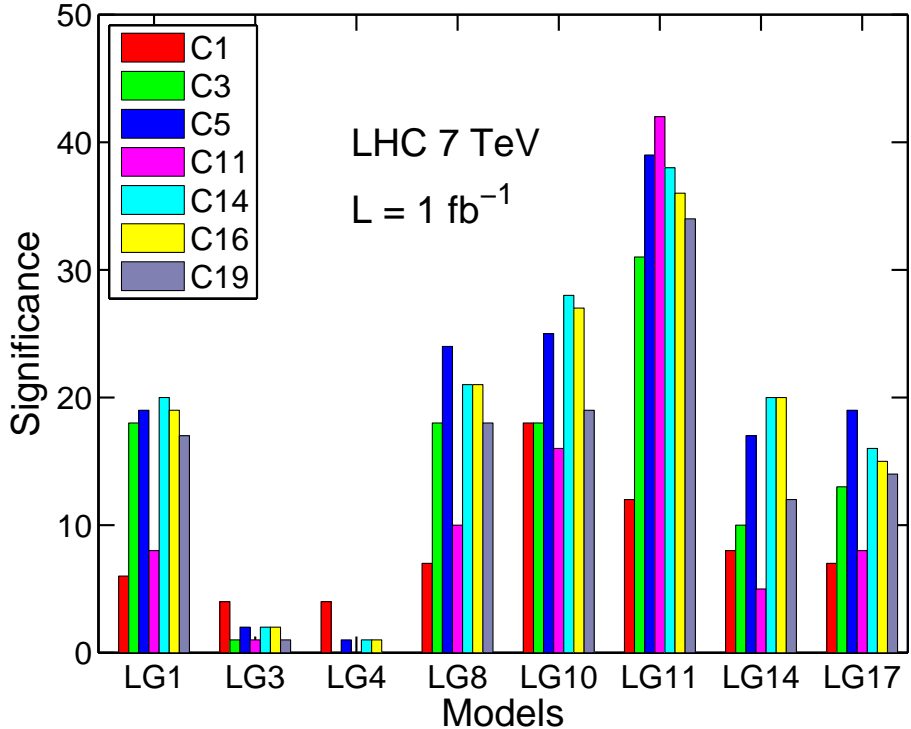}
\caption{(Color online)
A summary of the significance (SUSY signal divided by the square root of the SM background) for various signature channels$/$cuts  ($C$) 
for a subset of the low gluino (LG) mass  models
with an integrated luminosity  of  $1$ fb$^{-1}$ at the LHC for  $\sqrt s=7$ TeV. 
We will discuss in detail in Sec.(IV) how these drastic variations come about even 
though the gluino masses here are confined to the range $\sim$ (350-700) GeV for all models.   
}
\label{fig:significance}
\end{center}
\end{figure}

We give a broad sample of models in  Table(\ref{bench}),
where the SUGRA model parameters at the GUT scale are exhibited. The parameters include 
the universal scalar mass $m_0$, the three gaugino masses $M_{1,2,3}$, the universal trilinear 
coupling $A_0$, and the ratio of the two Higgs vacuum expectation values, $\tan\beta$. 
These models satisfy the stringent  Fermi-LAT constraints on 
$\gamma \gamma$ and $\gamma Z$ cross sections as discussed later.
The sparticle mass hierarchies corresponding to these models are given in Table(\ref{patterns}).  In Fig.(\ref{fig:significance}) we give a display of the significance for several optimal channels for
a sample of the  model classes at 1~fb$^{-1}$ of integrated luminosity  at $\sqrt{s} = 7~\rm TeV$.   One finds  that all of the models shown  are visible in several channel except for 
LG3 and LG4,  which are visible only in a small number of channels. We discuss these results in depth in Sec.(IV).  A more detailed discussion of these model  classes follows.

{\it 1. Low mass gluinos in mSUGRA: }
The possibility that  a low mass gluino  and heavy scalars can arise  in the radiative breaking of the electroweak symmetry (REWSB)
in SUGRA models was seen early on  \cite{ArnowittNath}. 
 It was later realized that this phenomenon 
 is more general, and the Hyperbolic Branch  (HB) or the  Focus Point (FP) region  of REWSB
 was discovered where scalars are heavy and gauginos are light
   \cite{Chan:1997bi}. 
 In the analysis  of  \cite{land2}, it was found that on the HB the chargino is predominantly the NLSP over a very broad class of soft breaking models
  \footnote{ HB models with larger Higgsino components and 
  with a light gluino have been studied in  \cite{Chan:1997bi,Baer:2007ya,FLNprl08},\cite{land1,land2,land3,GNLSP}, and Ref. 1 of \cite{recentBWE}.}.   
  
  \begin{table}[h]
\centering 
 Low mass gluino models in minimal SUGRA \\
\vspace{.3cm}
\begin{tabular}{|c||x{0.9cm}|x{0.9cm}|x{0.9cm}|x{0.9cm}|x{0.9cm}|x{0.9cm}|x{0.9cm}|x{0.9cm}||x{1cm}|x{1cm}|x{1cm}|x{1cm}|c|} 
\hline
Label	& $M_h$ & $M_{\na}$	  & $M_{\cha}$ & $M_{\tilde{g}}$ 
 &$M_{\tilde{\tau}_1}$ & $M_{\tilde{t}_1}$ & $M_A$& $M_{\tilde{b}_1}$ 
 & $Z_{11}$ & $Z_{12}$ & $Z_{13}$ & $Z_{14}$  
& $\sigma^{\rm SI}_{\na p} ({\rm cm}^2) $ \tnhl\hline
LG1 &  115 & 53 & 104 & 424 & 1983 & 1118 & 2040 & 1610 & 0.993 & -0.051 & 0.103 &  -0.025 & $5.4 \times 10^{-46}$\tnhl
LG2&111&118&220&715&126&476&463&606&0.990&-0.044&0.127&-0.051&$5.2\times10^{-45}$\tnhl
\end{tabular}
 \caption{mSUGRA models of a low mass gluino and a display of  some of the lighter masses within the sparticle mass hierarchies. In model LG1  the  neutralino, the chargino and the gluino are all light while the SUSY scalars are  heavy. This model  generates the relic density in the WMAP band via the annihilation of the neutralinos near the 
 Higgs pole. However, in model LG2 the scalars are also light and the relic density lies in the WMAP band 
  via   coannihilations of the neutralinos with the stau and other light sleptons.   All masses are in GeV.}
 \label{msp4}
\end{table}

  However, a low mass gluino, around 400 GeV, can arise in minimal supergravity if the  neutralinos annihilate near the Z-pole or near the  Higgs poles \cite{Nath:1992ty}. 
 Specifically a low mass gluino arises in mSUGRA in the sparticle mass landscape 
  which was labeled as mSP4 in   \cite{land1} where the first four sparticles
  have the mass hierarchy  as follows:
  \be 
~~{\rm Model~LG1: ~~} \na < \cha < \nb < \g, ~~~ M_{\g} = 424~\GeV~.
  \ee  
An illustration of such a model is given by LG1 in Table(\ref{msp4}). In this case the LSP is essentially dominated by the bino component\footnote{We use
the notation  $Z_{11},Z_{12},Z_{13},Z_{14}$ for  the eigen components of the lightest neutralino, namely the 
bino, the wino, and the  two  Higgsinos.
} as can be seen from Table(\ref{msp4}).
The model LG1 has a low mass gluino of  about 420~GeV, and the relic density is within the WMAP band 
due to neutralino annihilations near the Higgs pole. 
It is easily seen that the model LG1 is on the HB  and, thus, has heavy scalars. Since the neutralino is 
bino-like this leads to scaling of the gaugino mass spectra   \cite{Arnowitt:1992aq}. Further, because the neutralino has 
a suppressed Higgsino content, the spin-independent scattering is  suppressed at the level of 
$\sim 5 \times 10^{-46} \rm cm^2$.

A diametrical situation holds for the model point LG2. While the gluino is still light in this model, 
lying in the sub-TeV region, it is actually the heaviest sparticle in the whole sparticle spectrum.
Specifically the hierarchy is 
  \be 
~~{\rm Model~LG2: ~~} \na < \sta < \tilde{\ell}_R < \snl < \ldots <  \g, ~~~ M_{\g} = 715~\GeV.
  \ee     
The relic density in the model LG2 is satisfied via coannihilations \cite{GS} with the stau
 and with the other sleptons. 
Since model LG2 has a lighter scalar sparticle spectrum, 
it has a larger spin-independent cross section than LG1 by a factor of about 10.
The
ordering of the gluino mass in the hierarchy will 
have dramatic effects on the signals  at the LHC which we will discuss
in detail shortly.

  {\it 2. Low Mass Gluinos in Nonuniversal SUGRA models:}
 Low mass gluinos can arise from models where the gaugino masses are in general nonuniversal at the
 GUT scale, as for instance, from a combination of singlet and non-singlet 
 F term breaking  as discussed in the beginning of this section. Models LG3-LG17  in Table(\ref{bench})   fall in this class.  
Such models can lead to a low mass gluino that can be the NLSP,  the NNLSP, etc.  
 The case  when the gluino is the NLSP requires 
special attention since here the relic density can be satisfied by
the neutralino coannihilations with the gluino, as in the model LG3. This case was 
discussed at length in   \cite{Profumo2004,GNLSP,Wells}. For the models discussed in Table(\ref{gnlsp}), which have
a gluino NLSP (GNLSP) with a rather low mass gluino (i.e., below  400 GeV), the sparticle mass
hierarchies  are given by \cite{GNLSP}
  \be
~~{\rm Models~(LG3,LG4,LG5): ~~} \na < \g <   \cha  < \nb < \ta,~~~ M_{\g} = (386,378,385)~\GeV~. 
\ee

\begin{table}[t!]
\centering
 Low mass gluinos in GNLSP models\\
\vspace{0.3cm}
\begin{tabular}{|c||x{0.9cm}|x{0.9cm}|x{0.9cm}|x{0.9cm}|x{0.9cm}|x{0.9cm}|x{0.9cm}|x{0.9cm}||x{1cm}|x{1cm}|x{1cm}|x{1cm}|c|} 
\hline
Label	& $M_h$ & $M_{\na}$	  & $M_{\cha}$ & $M_{\tilde{g}}$ 
 &$M_{\tilde{\tau}_1}$ & $M_{\tilde{t}_1}$ & $M_A$& $M_{\tilde{b}_1}$ 
 & $Z_{11}$ & $Z_{12}$ & $Z_{13}$ & $Z_{14}$  
& $\sigma^{\rm SI}_{\na p} ({\rm cm}^2) $ \tnhl\hline
LG3&112&340&429&386&1253&455&1421& 995&0.997&-0.026&0.067&-0.029&$6.3\times10^{-46}$\tnhl
LG4&125&377&454&378&3529&1244&3888&2615&0.999& -0.006&0.021& -0.005&$8.4\times10^{-48}$\tnhl
LG5&117&365&660&385&1081&679&1167&1321 &0.999& -0.003&0.039&-0.012&$2.1\times10^{-46}$\tnhl
\end{tabular}
  \caption{A display of the lighter sparticle  masses within the mass hierarchies, 
    and other attributes of GNLSP models with low mass gluinos.  The mass splitting between the
gluino and neutralino is between $\sim (1 - 50) \rm GeV$
for these models. Further details are given in the text.   All masses are in GeV.}
 \label{gnlsp}
\end{table}

Since the gluino coannihilation in the 
 GNLSP model is a relatively new entry among the ways dark matter originates  in the early Universe, 
 we summarize the main features of the relic density calculation here first, before discussing the
 relevant features of this class  of models that enter in the LHC signature analysis. 
 Thus, for 
 a GNLSP the relic density depends strongly on coannihilation effects which are controlled by the Boltzmann factor \cite{GS} 
\be 
\gamma_i=\frac{n_i^{\rm eq}}{n^{\rm eq}} =
\frac{g_i(1+\Delta_i)^{3/2} e^{-\Delta_i x}} {\sum_j g_j
(1+\Delta_j)^{3/2}e^{-\Delta_j x}}~, 
\label{coann}
\ee
where $g_i$ are the
degrees of freedom of $\chi_i$, $x={m_1}/{T}$,
and $\Delta_i =(m_i-m_1)/m_1$, with $m_1$ defined as the 
LSP mass.  Thus, for the analysis of the relic density, 
the effective annihilation cross section $\sigma_{\rm eff}$ can be written approximately as  
\be
\sigma_{\rm eff} = \sum_{i,j} \gamma_i \gamma_j \sigma_{ij}
\simeq \sigma_{\g \g} \gamma^2_{\g}+2\sigma_{\g \na} \gamma_{\g}\gamma_{\na}
+\sigma_{\na \na} \gamma^2_{\na}\simeq  \sigma_{\g \g} \gamma^2_{\g}~,
\ee
where we have used the fact  that the gluino annihilation cross sections are usually much larger than the 
LSP annihilation even with inclusion of the Boltzmann factor
and 
\beqn
\sigma(\g\g\to q\bar q)&=&  {\cal E}_q {\pi\alpha_s^2\bar\beta\over 16\beta s}
(3-\beta^2)(3-\bar\beta^2)~,
\\
\sigma(\g\g \to gg)&=&{\cal E}_g{3\pi\alpha_s^2\over 16\beta^2s}\left\{\log{1+\beta\over
1-\beta}\left[21-6\beta^2-3\beta^4\right]-33\beta+17\beta^3\right\}~,
\label{xsecgg}
\eeqn
where the non-perturbative corrections to the annihilation cross section can arise via multiple gluon 
exchange, giving rise to a Sommerfeld enhancement factor $\cal E$. These effects may be 
approximated by \cite{Baer:1998pg}
\beqn
{\cal E}_j = {C_j \pi \alpha_s \over \beta} 
\left[ 1- 
exp \left \{-{C_j \pi \alpha_s\over \beta}\right\}
\right]^{-1}~, 
\eeqn
where $C_{j=g}= 1/2 (C_{j=q}= 3/2)$ for $\tilde g \tilde g \to g g (\tilde g \tilde  g \to q\bar q)$.
In the above  $\beta = \sqrt{1-4 m^2_{\g}/s}$,  and $\bar\beta=\sqrt{1-4m_q^2/s}$.  Although the gluino annihilation cross section 
$\sigma_{\g\g}$ varies with gluino mass, 
the Boltzmann suppression factor $\gamma_{\g}$  
controls the contribution to the $\sigma_{\rm eff}$ so that 
the relic density of the  bino-like neutralino is consistent with WMAP. 
We note that the effects of the Sommerfeld enhancement 
on the 
gluino cross sections can increase $\Delta_{\g}$ 
by a small amount for a bino-like LSP $\sim(2-3)\%$ \cite{GNLSP,Profumo2004}. 
Such an increase in the mass gap between the gluino and the neutralino 
can potentially enhance the discovery reach of this class of models at the  LHC as the mass gap between the $\g$ and $\na$ plays a crucial role in the strength of the LHC signals \cite{GNLSP} which we will discuss in details later.
 Three GNLSP models are exhibited in  Table(\ref{gnlsp}).  
 Some of their pertinent spectra and other attributes, including their spin-independent cross sections, are 
also given in Table(\ref{gnlsp}). 
It is seen from this table the neutralino is dominantly a bino. For each
of these models gluino coannihilation dominates the relic
density calculations. In the absence of additional hidden
sector gauge groups (to be discussed in what follows),
only LG3 lies in  the WMAP band, while models LG4 and LG5
each have a reduced relic abundance and reduced mass gaps between $\g$ and $\na$.

Next we consider  five models, LG6--LG10, in Table(\ref{bench}) where the gluino is 
light and in some cases it is  the next to next to LSP (GNNLSP). Specifically the sparticle mass
hierarchies are
  \beqn
{\rm Model~LG6: ~~} &\na < \ta < \g  < \cha < \nb, & M_{\g} = 391~\GeV~, \non
{\rm Model~LG7: ~~} &~~~ \na < \cha < \nb < \g  < \nc, & M_{\g} = 442~\GeV~, \non
{\rm Model~LG8: ~~}  &\na < \cha < \nb < \g  < \nc, & M_{\g} = 417~\GeV~, \non
{\rm Model~LG9: ~~} &\na < \cha < \nb < \n34 <  \chb  < \g, & M_{\g} = 696~\GeV~, \non
{\rm Model~LG10: ~~} & \na < \cha < \nb < \g <\nc <\ta, &  M_{\g} = 365~\GeV ~.
\eeqn
Their  relevant sparticle spectra and
other attributes including eigencontent, and spin-independent cross sections are displayed in Table(\ref{gnnlsp}).  Here the NLSP can be the stop
or the chargino, while the
 LSP is a mixture of bino, wino and Higgsino. The specific nature of the neutralino
eigencontent makes these models significantly different from each other and from other light
gluino models. This includes an interesting subclass of models where the LSP is dominantly Higgsino  (LG10).
Such a model class was  recently analyzed \cite{Chen:2010yi} in the context of both the PAMELA positron
excess and the Fermi-LAT photon line data.

\begin{table}[t!]
\centering
 Low mass gluinos including GNNLSP models\\
\vspace{0.3cm}
\begin{tabular}{|c||x{0.9cm}|x{0.9cm}|x{0.9cm}|x{0.9cm}|x{0.9cm}|x{0.9cm}|x{0.9cm}|x{0.9cm}||x{1cm}|x{1cm}|x{1cm}|x{1cm}|c|} 
\hline
Label	& $M_h$ & $M_{\na}$	  & $M_{\cha}$ & $M_{\tilde{g}}$ 
 &$M_{\tilde{\tau}_1}$ & $M_{\tilde{t}_1}$ & $M_A$& $M_{\tilde{b}_1}$ 
 & $Z_{11}$ & $Z_{12}$ & $Z_{13}$ & $Z_{14}$  
& $\sigma^{\rm SI}_{\na p} ({\rm cm}^2) $ \tnhl\hline
LG6&109&359&570&391&737&378&839& 833&0.992&-0.020&0.109&-0.064&$6.6\times10^{-45}$\tnhl
LG7&119&108&120&442&2786&1357&2947& 2185 &0.998&-0.049&0.039&-0.006&$2.4\times10^{-47}$\tnhl
LG8&112&104&117&417&1255&718&1284&  1032 &0.968&-0.202&0.143&-0.041&$2.7\times10^{-45}$\tnhl
LG9&115&135&152&696&1811&1067&1874& 1513&0.985&-0.136&0.098&-0.030&$7.8\times10^{-46}$\tnhl
LG10 &  112 & 111 & 115 & 365 & 1570 & 734 & 1609 & 1323 & 0.058 & -0.075 & 0.721 &  -0.686 & $7.3 \times 10^{-45}$\tnhl
   
\end{tabular}
 \caption{The spectrum of low mass sparticles including the GNNLSP models  within the sparticle mass hierarchies, 
  and other attributes of models in NUSUGRA with a low mass gluino. Model LG6 is a GNNLSP,
 and models LG7,LG8, and LG9 are effectively GNNLSP as the chargino and second heaviest neutralino are roughly mass degenerate. LG10
 has a mass splitting  between the chargino and the neutralino of $\sim 5 $ GeV and  is  effectively a GNNLSP model.
  All masses are in GeV.}
  \label{gnnlsp}
\end{table}

\begin{table}[t!]
\centering
{Models with a  low mass gluino with a light stau and a light stop}\\
\vspace{0.3cm}
\begin{tabular}{|c||x{0.9cm}|x{0.9cm}|x{0.9cm}|x{0.9cm}|x{0.9cm}|x{0.9cm}|x{0.9cm}|x{0.9cm}||x{1cm}|x{1cm}|x{1cm}|x{1cm}|c|} 
\hline
Label	& $M_h$ & $M_{\na}$	  & $M_{\cha}$ & $M_{\tilde{g}}$ 
 &$M_{\tilde{\tau}_1}$ & $M_{\tilde{t}_1}$ & $M_A$& $M_{\tilde{b}_1}$ 
 & $Z_{11}$ & $Z_{12}$ & $Z_{13}$ & $Z_{14}$  
& $\sigma^{\rm SI}_{\na p} ({\rm cm}^2) $ \tnhl\hline
LG11&103&121&134&433&516&271&701&489&0.978&-0.171&0.117&-0.032&$6.6\times10^{-45}$\tnhl
LG12&107&169&190&588&424&489&551&602&0.973&-0.171&0.146&-0.058&$3.0\times10^{-44}$\tnhl
LG13&104&160&359&684&365&187&845&604&0.996&-0.023&0.076&-0.041&$2.0\times10^{-45}$\tnhl
LG14&111&114&227&618&118&338&478&514&0.990&-0.043&0.124&-0.052&$4.7\times10^{-45}$\tnhl
LG15&109&121&238&602&132&397&398&525&0.979&-0.052&0.177&-0.083&$1.7\times10^{-44}$\tnhl
\end{tabular}
\caption{An exhibition  of the light sparticle masses within the hierarchies in models with low mass gluinos that are accompanied by light staus 
 and light stops. The light Higgs masses for these models lie in the range 
 (103-111) GeV and are generally lighter than the models where the chargino or the gluino is the NLSP shown in the previous tables, as
 the scalars in such model are much heavier. All masses are in GeV. }
 \label{nonsinglet}
\end{table}

Finally, in Table(\ref{nonsinglet}) we  give five models, LG11--LG15 where in addition to the low mass gluino one also has a light
stau and a light stop (due to the smaller GUT value of $m_0$) along with  a light chargino.  
Models LG11--LG15 have compressed 
sparticle spectra with the heaviest  sparticle mass around 850~GeV. Model LG11 has a highly 
reduced overall mass scale of the sparticles, with a low mass gluino and a  light stop
with the mass hierarchy
  \be
~~{\rm Model~LG11: ~~} \na <  \cha < \nb < \ta < \g,~~~ M_{\g} = 433~\GeV~, 
\ee
with the remaining sparticles in the mass range (500-700)~GeV.
 In each of these models the relic  density 
 can be satisfied via coannihilations with different superparticles,
and, in particular, model LG13 proceeds via stop coannihilations.
For LG12, the gluino mass lies 
in the middle of the sparticle mass spectra, while for  (LG13-LG15) the gluino mass is close to being the
largest mass even though it is still relatively light, i.e.,  $M_{\tilde g} < 700 ~\rm GeV$. 
Models LG11 and LG13 have rather low-lying light Higgs. However, the extraction of the Higgs mass from
LEP data is model dependent, and  we retain these  models in the analysis pending further experimental data.

 {\it 3. Low mass gluinos in extended SUGRA models and PAMELA data:} Recently, 
the PAMELA collaboration \cite{PAMELA} has included an analysis of statistical uncertainties in the positron fraction  \cite{statistical}
and presented its results on the absolute $\bar p $ flux  \cite{PAMELApbar}.
Models LG10, LG16, and LG17 
in Table(\ref{bench}) have  low mass gluinos with many other desirable
  features.  Specifically, they can explain the positron excess in the PAMELA satellite experiment  
  \cite{PAMELA}
  (for  previous experiments see  \cite{heatams}).
  However, their relic density by the usual thermal annihilation processes is rather small.
  This situation can be modified at least in two ways. 
  
    \begin{table}[t!]
\centering
 Low mass gluino SUGRA models which explain the PAMELA positron excess\\
\vspace{0.3cm}
\begin{tabular}{|c||x{0.9cm}|x{0.9cm}|x{0.9cm}|x{0.9cm}|x{0.9cm}|x{0.9cm}|x{0.9cm}|x{0.9cm}||x{1cm}|x{1cm}|x{1cm}|x{1cm}|c|} 
\hline
Label	& $M_h$ & $M_{\na}$	  & $M_{\cha}$ & $M_{\tilde{g}}$ 
 &$M_{\tilde{\tau}_1}$ & $M_{\tilde{t}_1}$ & $M_A$& $M_{\tilde{b}_1}$ 
 & $Z_{11}$ & $Z_{12}$ & $Z_{13}$ & $Z_{14}$  
& $\sigma^{\rm SI}_{\na p} ({\rm cm}^2) $ \tnhl\hline
LG10 &  112 & 111 & 115 & 365 & 1570 & 734 & 1609 & 1323 & 0.058 & -0.075 & 0.721 &  -0.686 & $7.3 \times 10^{-45}$\tnhl
LG16&112&182&188&343&2193&1254&2290&1780&0.654&-0.718&0.209&-0.112&$3.0\times 10^{-44}$\tnhl
LG17&111&168&171&425&2986&1698&3215&2421&0.724&-0.681&0.101&-0.043&$5.9\times10^{-45}$\tnhl
\end{tabular}
 \caption{An exhibition of the sparticle  mass hierarchies in low mass gluino models  which can explain the PAMELA positron excess. 
The models are consistent with data from the 
Fermi-LAT  \cite{linesource},  
CDMS-II  \cite{CDMS}, 
XENON100  \cite{Aprile:2010um}, and EDELWEISS-2 \cite{collaboration:2010ei}
experiments. In the above we do not list the component of the
neutralino lying in the hidden sector as it  is typically small, i.e., $Z_{1k}^h <1\%$.
 }
\label{pamela}
\end{table}

First, the relic density could be enhanced via coannihilation 
  with matter in the hidden sector.  For example, an extended Abelian  gauge symmetry can arise from the hidden sector and can couple to the 
 MSSM sector via  mass mixing or  kinetic mixing with the hypercharge field. The mass mixing arises
 via the \st mechanism and  the kinetic mixing via mixing  of the hidden sector 
 field strength and the hypercharge field strength.
  Such mixings can lead to an enhancement of the relic density which we now discuss.  Thus,  we consider
  for specificity
   the mass  growth via  the Stueckelberg mechanism 
allowing for mixings between the hypercharge
gauge multiplet ($Y_{\mu}, \lambda_Y, \bar\lambda_Y, D_Y$) and the gauge multiplet of $U(1)_X$ 
($X_{\mu}, \lambda_X, \bar \lambda_X, D_X$) both taken in the Wess-Zumino gauge. The effective Lagrangian contains the mixing terms  
\cite{korsnath}, Ref (1-4) of \cite{fln1}, and Ref. \cite{fln2}
\be
 -\frac{1}{2} (\partial_{\mu} \sigma + M_Y Y_{\mu} +M_X X_{\mu})^2 +[\psi_{\rm st} (M_X \lambda_X +M_Y \lambda_Y) + h.c. ]~,  
\label{st1}
 \ee
 where 
 the axionic field $\sigma$ arises from a 
 chiral multiplet  $S =(\rho+ i\sigma, \psi_{\rm st},F_S)$.
The fields $\psi_{\rm st}$ and $\lambda_X$  produce  two Majorana spinors 
(hidden sector neutralinos) which mix with the neutralinos in the visible sector via the 
mass mixing given by Eq.(\ref{st1}). 
In addition the 
 hidden sector neutralinos can mix with the visible sector via kinetic mixing \cite{fln2,Dienes:1996zr}
\be
 -\frac{\delta}{2} X_{\mu \nu} Y^{\mu \nu} -i\delta(\lambda_X \sigma \cdot \partial \bar \lambda_Y+ (Y\leftrightarrow X) ) +\delta D_X D_Y~. \ee
We note that $M_Y:M_X$ and $\delta$ are constrained to be small  by fits to the precision electroweak data \cite{fln1}
and thus the additional neutralino states are weakly coupled to the MSSM. 

An estimate of the bound on the (mass or kinetic) mixing can be 
obtained from studying eigenvalues of the vector sector. One finds the mixing is constrained for
masses near the Z-pole   as derived in Ref. 1 of \cite{fln1}
\be 
|\epsilon| \lesssim .05  \sqrt{1-m^2_1/m^2_2}~,
\ee
where $\epsilon$ is the overall mass and/or kinetic mixing and  $m_{(1,2)}$ is the mass of the hidden sector vector boson, or the $Z$ boson mass, depending
on which side of the Z-pole the hidden sector mass resides. 
We add that dark matter with Stueckelberg mass growth and/or kinetic
mixing, with a massive hidden $U(1)$ \cite{fln2}
has also appeared in Refs. \cite{Fucito:2008ai} in  various  contexts.

The analysis above can be extended to a $U(1)^n_X$ gauge group in the hidden sector that produces
$2n$ additional Majorana fields in the hidden sector, which we denote collectively by $\xi^h_k$ , $(k=1-2n)$. 
This extended set of hidden sector Majoranas will mix with the visible sector via mass mixing and kinetic
mixing, where we again  take the mixings  to be small. 
Due to the small mixings between the visible and the hidden sector Majoranas, the LSP in the visible sector will  have a small component which lies in the hidden sector, and
as such, the eigencontent of the LSP is modified  so that \cite{korsnath,fln2,FLNN,Arvanitaki:2009hb} 
\beqn
\tilde\chi_1^0=Z_{11} \tilde B + Z_{12} \tilde W + Z_{13} \tilde H_1 + Z_{14} \tilde H_2
+ \sum^{2n}_{k=1}Z_{1k}^{h} \xi_k^{h}~, 
\label{eigen2}
\eeqn
where $\xi_{k}^{h}$ are  the hidden sector neutralinos (Stinos)
 composed of the hidden sector gauginos and the hidden sector chiral fields as  discussed above and $Z_{1k}^{h}$ 
record the leakage into the hidden sector. Because of the small mixings between the visible
and the hidden sector, the $Z_{1k}^{h}$ are rather small, typically less than $1\%$ of the 
components in the visible sector. 
For this reason we do not record them in 
Table(\ref{pamela}) and elsewhere.  However,  we will assume that the overall
mixing is large enough so that the extra-weakly interacting states \cite{fln2,hall}
remain in contact with the thermal bath. Specifically, we envision mixings
in the range $(10^{-5} -10^{-2}).$  In what follows, in the context of the PAMELA
data, we will assume that the hidden sector states lie above the Z-pole. The system of additional
Majoranas  then mostly have hidden sector eigencontent and have masses close
to the mass of the LSP.  

 \begin{figure*}[h!!!]
  \includegraphics[width=11cm,height=8cm]{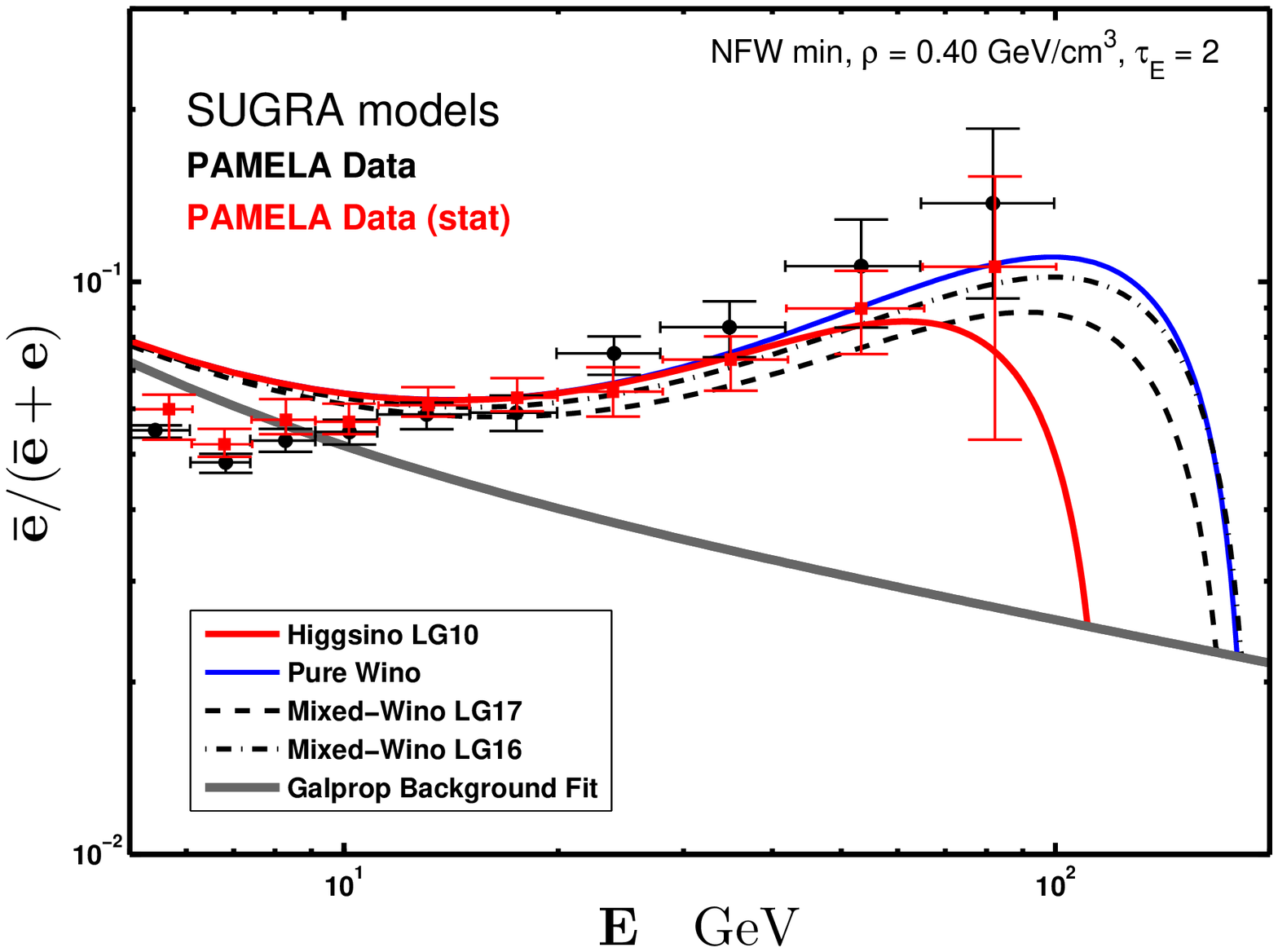} 
  \includegraphics[width=11cm,height=8cm]{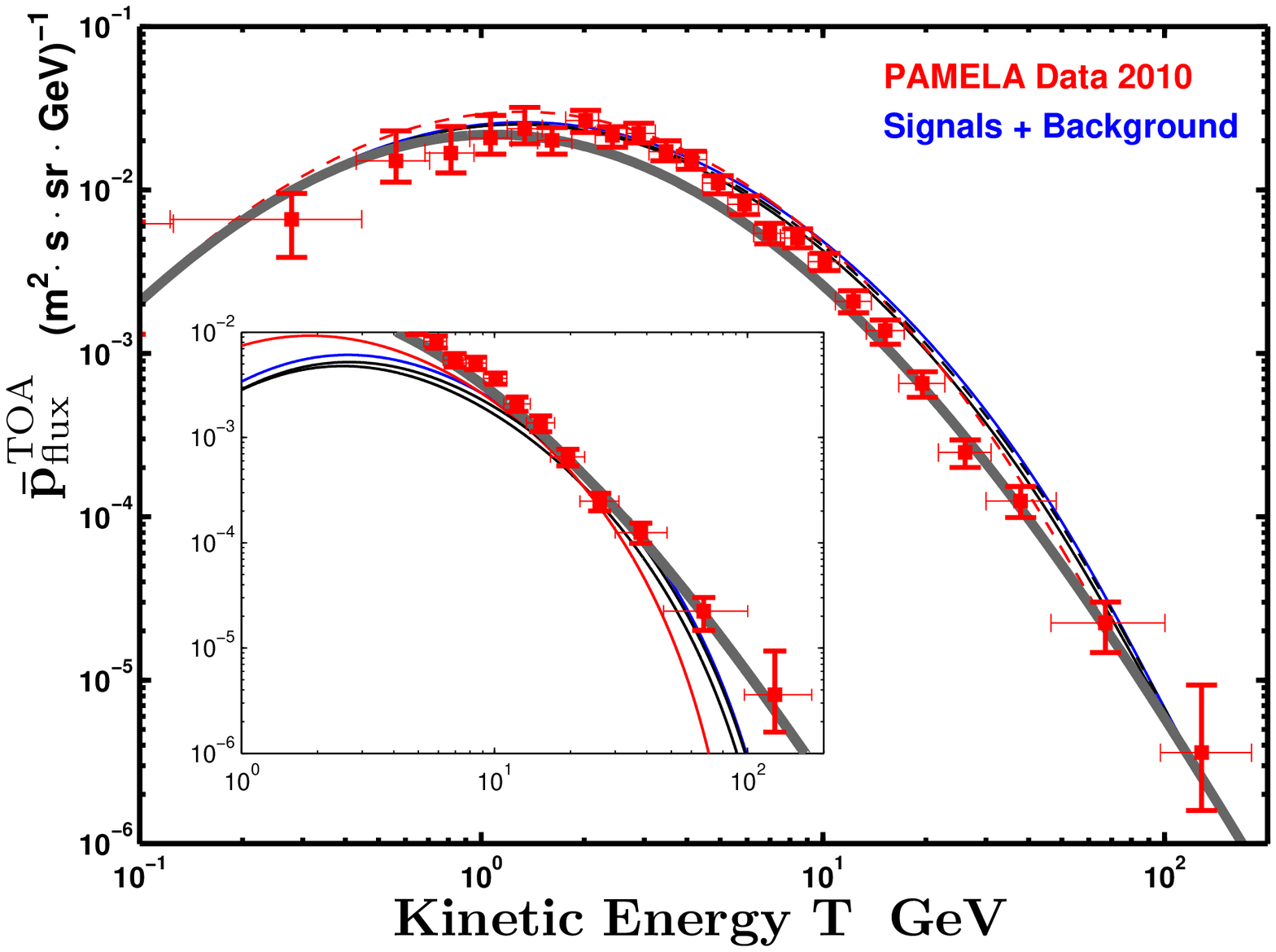}  
\caption{(Color online) Top:  Mixed-wino LSP models LG16  and LG17, the Higgsino LSP model LG10,  and the PAMELA positron excess.  For comparison
a pure wino LSP model is  also shown. Boost factors of (1-3) are used. We note that an LSP with 
a pure wino content while compatible with the PAMELA data has difficulties with the 
$\gamma\gamma$ and $\gamma Z$ cross sections from Fermi monochromatic photon data.
 For this reason Table(\ref{bench}) does not contain a pure wino model.
 Bottom: The PAMELA anti-proton flux and the SUSY models. The $\bar p$ flux is compatible
 with the data. The mixed-wino  models as well as the Higgsino model  can produce large LHC jet signatures  as discussed in the text.  
  }
\label{cosmic}
\end{figure*}   

We consider now  the inclusion of the hidden sector 
which may have a  very significant effect on the relic density.
For  the  case that the Majoranas in the hidden sector
 are effectively degenerate in mass with the LSP,
    the  LSP  can coannihilate with the hidden sector neutralinos generating an enhancement for
   the  density of relic neutralinos relative to that in the MSSM, through the extra degrees
   of freedom supplied by the hidden sector, so that
   \be
 \Omega_{\tilde{ \chi}^0_{1}} h^2  \simeq ~f_{E}  \times \Omega_{\tilde \chi_1^0}^{\rm MSSM}  h^2~, 
 \label{enhance}
  \ee  
 where $f_E$ is the relic density enhancement factor  given by \cite{FLNN}
  \be
f_E= \left[1 + \frac{d_{\rm hid}}{d_{\rm vis}}\right]^2~,
 \ee
and  where $d_{\rm hid}$ is the degree of degeneracy in the hidden sector and $d_{\rm vis}$ is 
the degree of degeneracy in the visible sector.
We note in passing that an analysis on the effects of fields  in the hidden sector on
the relic density via coannihilation effects was studied in \cite{fln2} and a similar idea was
pursued in \cite{Profumo:2006bx} but without an additional hidden sector.  For the $U(1)^n_X$  extension of the SUGRA 
model with the hidden sector neutralinos essentially degenerate in mass with the LSP, one has 
$d_{\rm hid} = 2\times 2n$, where $2n$ is the number of Majorana  fields  in the hidden sector with
2 degrees of freedom for each Majorana field. For the case when the LSP neutralino is pure wino, the chargino is essentially
degenerate in mass   with the LSP and in this case 
$d_{\rm vis}=2{(\rm LSP)} + 4{\rm (chargino)}$, 
while for the case when the LSP neutralino is a mixture of bino, wino, and Higgsinos, one has
$d_{\rm vis} \to 2{(\rm LSP)}$.  Thus, for the case when the neutralino and the chargino are not
degenerate as is the case when the eigencontent of the neutralino is a mixture of bino, wino
and Higgsino,  a large $f_E$ can be generated, for example,  
$f_E=49$ for $n=3$, where the maximal $f_E$ we obtain is $\sim 40$ due to coannihilations in the visible sector.  We will utilize this enhancement 
of the relic abundance  in fitting the WMAP data. 

\begin{table}[t!]
\begin{center}
$\left<\sigma v \right>^{\text{Theory}}_{\gamma Z}$ and $\left<\sigma v \right>^{\text{Theory}}_{\gamma \gamma}$ \\
\vspace{0.3cm}
\begin{tabular}{|c||c|c||c|c|}
\hline
Label	&	$E_{\gamma}$ (GeV)	&	$\left<\sigma v \right>^{\text{Theory}}_{\gamma Z}$(cm$^3$/s)	&	$E_{\gamma}$ (GeV)	&	$\left<\sigma v \right>^{\text{Theory}}_{\gamma \gamma}$ (cm$^3$/s)			\tnhl
LG1	&	14	&	$2.7\times 10 ^{-30}$	&	53	&	$1.5\times 10^{-34}$			\tnhl
LG2	&	101	&	$2.6\times 10^{-30}$	&	118	&	$9.4\times 10^{-30}$			\tnhl
LG3	&	334	&	$1.7\times10^{-32}$	&	340	&	$2.5\times 10^{-32}$			\tnhl
LG4	&	372	&	$2.2\times 10^{-34}$	&	377	&	$6.9\times10^{-34}$			\tnhl
LG5	&	359	&	$1.1\times 10^{-33}$	&	365	&	$9.1\times 10^{-33}$			\tnhl
LG6	&	353	&	$3.2\times 10^{-32}$	&	359	&	$6.7\times 10^{-32}$			\tnhl
LG7	&	89	&	$1.2\times 10^{-28}$	&	108	&	$2.8\times10^{-29}$			\tnhl
LG8	&	84	&	$1.5\times 10^{-28}$	&	104	&	$3.5\times 10^{-29}$			\tnhl
LG9	&	119	&	$8.8\times 10^{-29}$	&	135	&	$1.7\times10^{-29}$			\tnhl
LG10	&	92	&	$1.8\times10^{-28}$	&	111	&	$8.7\times 10^{-29}$			\tnhl
LG11	&	103	&	$4.0\times 10^{-29}$	&	121	&	$7.1\times10^{-30}$			\tnhl
LG12	&	157	&	$3.5\times 10^{-29}$	&	169	&	$5.0\times 10^{-30}$			\tnhl
LG13	&	147	&	$1.3\times 10^{-31}$	&	160	&	$4.0\times 10^{-31}$			\tnhl
LG14	&	96	&	$3.1\times 10^{-30}$	&	114	&	$1.1\times 10^{-29}$			\tnhl
LG15	&	104	&	$3.1\times 10^{-30}$	&	121	&	$7.9\times 10^{-30}$			\tnhl
LG16	&	171	&	$6.0\times 10^{-27}$	&	182	&	$1.1\times10^{-27}$			\tnhl
LG17	&	156	&	$9.4\times 10^{-27}$	&	168	&	$1.7\times 10^{-27}$			\tnhl
\end{tabular}
\end{center}
\begin{center}
\begin{tabular}{|c|c|c|c||c c|}
\hline
$E_\gamma$ (GeV)    & NFW & Einasto & Isothermal  & Model &   
$\langle \sigma v\rangle_{(\gamma Z),[\gamma\gamma]}^{\rm theory}$\\
\hline
(90--100)[110--120]   &  (6.0--3.8)[1.0--1.6] &  (4.3--2.8)[0.7--1.1] & (10.3--6.6)[1.7--2.7] &  LG10 & (0.18)[0.087] \\
(170--180)[180--190]  &  (4.0--6.1)[2.7--3.2] &  (2.9--4.4)[1.9--2.3]&  (6.8--10.4)[4.6--5.5] & LG16  & (6.0)[1.1]  \\
(150--160)[160--170]  &  (8.2--6.3)[2.7--1.7] &  (5.9--4.5)[2.0--1.3]  & (14.1--10.9)[4.7--3.0] & LG17 & (9.4)[1.7]\\ 
\hline
\end{tabular}
\caption{
Top table:  Predictions for $\tilde\chi_1^0 \tilde\chi_1^0 \to \gamma \gamma, \gamma Z$ for the models
 including the bino-like models, Higgsino-like models and wino-like models.
 Lower table:  The current experimental constraints from Fermi-LAT \cite{linesource} on 
$\gamma\gamma $ and $\gamma Z$ modes are shown with 
 three halo profiles.  Constraints are shown for those model classes which can describe the PAMELA
data, and have large neutralino self annihilation cross sections into $\gamma \gamma $ and $\gamma Z$. 
Shown are the ranges where the models can be constrained where the notation is  $[..]$  for the $\gamma \gamma$ mode and $(..)$
is for the $\gamma Z$ mode. All cross sections in the lower table are given in $10^{-27}$cm$^{3}$s$^{-1}$.
Models LG16, LG17 are within reach of the Fermi-LAT limits as they have a large wino content,
and are discussed in detail in the text.
}
\label{fermi}
\end{center}
\end{table}

With the above  enhancement mechanism one can achieve 
a large 
 $\langle \sigma v \rangle_{\rm halo}$ needed for a solution
to the PAMELA data and at the same time have a relic density close to the WMAP data.  This indeed is the case for models  LG16 and LG17 
which can explain the PAMELA data via the neutralino annihilation 
$\tilde\chi_1^0\tilde\chi_1^0\to W^+W^-$ with the
subsequent  dominant decay 
$W^+\to \ell^+ \nu_{\ell}$ while at the same time satisfying the relic density constraint via the additional 
hidden sector degrees of freedom. In these cases, the mixed-wino annihilation cross section into $WW \sim 6\times 10^{-25}~\rm cm^3s^{-1}$,
and only small effects from the possible clumps in the local halo are needed.
  For LG10, the Higgsino-like LSPs  annihilate into both $WW$ and $ZZ$,
where in addition   the decay  $Z\to  \ell^+ \ell^{-}$  gives a
contribution  to the positron excess. Here  one finds 
$\langle\sigma v\rangle_{WW+ZZ}\sim 3.5 \times 10^{-25}~\rm cm^3s^{-1}$ and one needs a clump factor
of around (2-3) to fit the PAMELA data while the LSP  mass is significantly lower, $\sim 111 \rm ~GeV$ \cite{Chen:2010yi}, than 
the mixed wino case. 
Interestingly, this Higgsino-like model produces  smaller $\bar p$ flux at higher energies and slightly larger flux at lower energies relative to the
 heavier mixed wino case. Also the Higgsino-like  
LSP receives less stringent constraints from the Fermi photon line data than the 
wino-like LSP.
We note that models with a Higgsino-like LSP have been studied in the past \cite{Kane:2001fz}.
 However, to our knowledge the 
 model class uncovered in  \cite{Chen:2010yi}, for which  LG10 is an example, is the first illustration of a SUGRA model with a Higgsino LSP that gives a fit  to the
PAMELA data.

To elaborate further,  the models of  Table(\ref{pamela})  have a neutralino mass in the range of 
$(111-183)$~GeV 
which allows a $\langle \sigma v \rangle_{\rm halo}$ 
in the range   $(3.5-6)~\times~10^{-25}$~$\rm cm^3s^{-1}$, and thus the model  can
produce the positron excess seen in the 
PAMELA experiment. 
 Further, as seen in Table(\ref{pamela}) these models 
are consistent with the current XENON100 
limits, and have the possibility to produce  scattering cross sections that are discoverable
in improved dark matter experiments, i.e., in the range $\sim 5 \times (10^{-45}-10^{-44})~\rm cm^2$. In addition, these models are consistent 
with the Fermi-LAT photon data as seen in Table(\ref{fermi}). 
Hence,  the eigencontent of the LSP in these models is
mixed-wino 
for LG16  and LG17, and close to being pure Higgsino for LG10, and therefore their 
dark matter signatures are modified drastically
relative to other models classes discussed which are significantly bino-like.

In the top  panel of Fig.(\ref{cosmic}) 
we show several fits to the PAMELA positron fraction where we include the data from recent experiments \cite{PAMELA,statistical}.  The analysis of this panel shows general agreement with  \cite{KLW,FLNN,Feldman:2010uv} for the pure or essentially
pure wino case
(for early work on cosmic rays relevant to this discussion see  \cite{Baltz}  and \cite{someprofiles}). Here, however, we specifically show that the mixed-wino models LG16 and LG17
and the Higgsino LSP model case, LG10,
provide a good description of the PAMELA data with boosts of order unity. For comparison we also
show the essentially pure wino case, which requires no boost, but has difficulty explaining the 
Fermi photon data.  
In the lower panel of Fig.(\ref{cosmic}) we  give a comparison of
the $\bar p $ flux with recently released data  \cite{PAMELApbar}. Indeed it is seen that the theoretical 
prediction of the
 $\bar p$ flux is in good agreement
with this  data.  The boost factors used are rather minimal and
 are not assumed different for the $\bar e$ and $\bar p$ fractions, which is in principle a possibility (different boosts are often
introduced in order not to upset the $\bar p$ flux while enhancing the $\bar e$ flux).
The analysis we present does not attempt to explain the 
ATIC/Fermi high energy $e+ \bar e$ data.  Such data could be explained  with  an additional electron source\cite{LR,KLW} with a wino LSP\cite{MoroiRandall,Kane:2001fz,KLW}
or mixed-wino LSP \cite{FLNN,Chen:2010yi}.

\section{IV: Signature Analysis at the LHC at $\sqrt s=7$ TeV \label{signatures}}
Signature analyses at center of mass energies of $\sqrt s=10$ TeV and $\sqrt s=14$ TeV at the LHC 
already
exist in the  literature, 
and as mentioned in the introduction, a few analyses at the
center of mass energies of $\sqrt s= 7$ TeV  have also appeared \cite{flnearly,Baer:2010tk,Peim,Feldman:2010uv,Giudice:2010wb,Wacker,moreearly,Chen:2010yi}. 
 For this analysis, our emphasis is on the discovery of models which admit low mass gluinos in early runs at the LHC consistent
with dark matter interpretations for a neutralino LSP. 

 {\it{ 1. Standard Model Background: }} The discovery of new physics requires an accurate determination of the standard model (SM) background.  The recent works 
 of   \cite{Baer:2010tk,Peim} have given an analysis of such backgrounds including $2\to n$ processes at $\sqrt s=7$ TeV
 appropriate for $pp$ collisions at the LHC.  We  use for our analysis the simulated SM background of   \cite{Peim} 
  which was generated with MadGraph 4.4~\cite{Alwall:2007st} for  parton level processes, Pythia~6.4~\cite{pythia} for  hadronization and PGS-4~\cite{pgs} for detector simulation.  An MLM matching algorithm with a $k_T$ jet clustering scheme was used to prevent double counting of final states. Further, the $b$-tagging efficiency in PGS-4 is based on the Technical Design Reports  of ATLAS~\cite{Aad:2009wy}
  (see \cite{Peim}), which is similar to the efficiency of CMS~\cite{Bayatian:2006zz}, with the mis-tagging rate of $b$-jet unmodified from the default in PGS-4.  In addition  Tauola is called for tau decays  \cite{tau}.  
   The processes that are included in the SM background are :
($\text{QCD } 2, 3, 4 \text{ jets}$),
($t\bar{t}+0,1,2\text{ jets}$),
($b\bar{b}+0,1,2\text{ jets}$),
($Z/\gamma\left(\to l \bar{l}, \nu \bar{\nu}\right)+0,1,2,3\text{ jets}$),
($W^{\pm}\left(\to l\nu\right)+0,1,2,3\text{ jets}$),
($Z/\gamma\left(\to l \bar{l}, \nu \bar{\nu}\right)+t\bar{t}+0,1,2\text{ jets}$),
($Z/\gamma\left(\to l \bar{l}, \nu \bar{\nu}\right)+b\bar{b}+0,1,2\text{ jets}$),
($W^{\pm}\left(\to l\nu\right)+b\bar{b}+0,1,2\text{ jets}$),
($W^{\pm}\left(\to l\nu\right)+t\bar{t}+0,1,2\text{ jets}$),
($W^{\pm}\left(\to l\nu\right)+t\bar{b}\left(\bar{t}b\right)+0,1,2\text{ jets}$),
($t\bar{t}t\bar{t}$,   $t\bar{t}b\bar{b}$,    $b\bar{b}b\bar{b}$),
($W^{\pm}\left(\to l \nu\right)+W^{\pm}\left(\to l \nu\right)$),
($W^{\pm}\left(\to l \nu\right)+Z\left(\to~all\right)$),
($Z\left(\to~all\right)+Z\left(\to~all\right)$), ($\gamma + 1, 2, 3 \text{ jets}$).
Here $l$ is $e,\mu,\tau$, a jet refers to gluon as well as first and second generation quarks and $all$ denotes either $l$, $\nu$ or jet.  The above processes are final state processes at the parton level, i.e. before hadronization.  
A more detailed discussion of the SM background can be found in \cite{Peim} which includes a list 
 of cross section, number of events and luminosity for each process (see Table(I) of \cite{Peim}).

 {\it 2. SUSY Signal Generation and Optimization of signature cuts: }  The  sparticle spectrum for the signal analysis was generated using SuSpect \cite{Djouadi:2002ze} via micrOMEGAs \cite{Belanger} and  branching ratios are computed with SUSY-HIT \cite{Djouadi:2006bz}.  
Some differences in the output of the sparticle masses are known to exist when computed with different codes.  
We have checked our models with SOFTSUSY~\cite{Allanach:2001kg} and found only small differences. 
 
 We now discuss a set of signatures used 
in our early discovery analysis at the LHC at $\sqrt s= 7$~TeV. 
The notation used in the these signatures  is:
   $\ell$ denotes $e, \mu$, and 
$p_T(\ell)$, $p_T(j)$  define the transverse momentum of the 
lepton $\ell$, and of the  jet $j$, while $n(\ell)$, $n(j)$ give us the
number of   leptons ($\ell$) and the number of jets ($j$) in the event. 
 We investigate a large number of cuts on the $ p_T$ of jets and leptons in
combination with  transverse sphericity, $S_{T}$, and missing energy, $\slashed{E}_{T}$.  For clarity,  we order objects by their $p_{T}$, i.e. the hardest jet would be denoted $j_{1},$ and we define $m_{\rm eff}$ and ${H}_T$ as  
\be
m_{\rm eff}=\displaystyle\sum_{i=1}^4 p_T\left(j_i\right)+\slashed{E}_T,~~~~~~ H_T = \displaystyle \sum_{i=1}^4 p_T(x_i)+\slashed{E}_T~,\label{meff}
\ee
where $x_i$ is a {\it visible} object and 
 where the sum is over the first four hardest objects
 \footnote{One could also define a $b$ jet effective mass to be the sum of the $p_T$s of the  four hardest $b$-jets.
 However, accuracy with which $p_T$ of the $b$ jets can be determined  may not  be high in early runs and thus the use of $b$-jet 
 effective mass may be experimentally challenging. Several alternate definitions of $H_T$ appear in the literature.}.
In this analysis  we define a signal that produces $S$ events to be discoverable for a particular signature cut if $S\geq\max\left\{5\sqrt{B},10\right\}$, where $B$ is the number of SM background events.  

In the analysis we investigate a broad set of cuts to enhance the significance.  The optimal cuts were found by varying the bounds on observables.  First, a broad optimization was carried out where the varied observables include  missing energy (100~GeV to 800~GeV in steps of 50~GeV), transverse sphericity  of all visible objects (with a lower bound of 0.15 to 0.25 
in steps of 0.05), number of jets (2 to 6 in integer steps), number of $b$-jets (0 to 3 in  integer steps), as well as the $p_T$ of the hardest jet (10~GeV to 500~GeV in steps of 100~GeV)  and second hardest jet (10~GeV to 250~GeV in steps of 50~GeV). 
Further, this optimization includes varying  the number of jets, the $p_T$ of the hardest jet, $p_T$ of the second hardest jet, and the $p_T$ of all the jets.  
  For the particular cuts, C17 and C18, that deal with opposite sign same flavor (OSSF) leptons, a Z-veto is applied, i.e. the leptons invariant mass  is not in the 76 GeV to 105 GeV region.  We note that in a preliminary scan, before the large optimization, a variation on cuts for  the  $p_{T}$ of leptons was also investigated. However,  it was
  found to be of little use in enhancing the significance  at low luminosity for the models we discuss.  Given this we omitted cuts on lepton $p_T$ from the large optimization.  
  However, the tri-leptonic signal  is an important signature for the discovery of supersymmetry via the 
 off-shell decay of the $W$ as well as other off-shell processes \cite{trilep}.  The tri-leptonic signal has been considered in the
 analyses at 7 TeV in recent works \cite{flnearly,Baer:2010tk}.
 A third optimization search was done in this channel by varying transverse sphericity (either no cut or $\geq 0.2$), missing energy ($\geq 100\GeV, 150\GeV, 200 \GeV, 250\GeV$), number of jets ($\geq 2,3,4,5,6$), $p_T(j_1)$ (no cut or $\geq60\GeV,100\GeV,150\GeV$) and $p_T(j_2)$ (no cut or $\geq20\GeV,30\GeV,40\GeV,60\GeV,80\GeV,100\GeV$). 
  
A subset of cuts found using the procedure above are listed below.    
   In choosing  these cuts 
     we have taken into account the uncertainty of how well missing energy can be determined in the early runs.  For this reason we have taken lower values of $\slashed{E}_T$, sometimes as low as $100\GeV$.  Better optimization can occur with other choices, specifically for larger values of $\slashed{E}_T$.  However, this requires a greater degree of confidence on how well $\slashed{E}_T$ is determined in the early runs.  

\begin{enumerate}

\item[C1:]

$\met100$~,

\item[C2:]

$\transph$, $\met100$~,

\item[C3:]

$\transph$, $\met100$, $n(\ell)=0$, $p_T(j_1)\geq 150\GeV$, $p_T(j_2,j_3,j_4)\geq40\GeV$~,

\item[C4:]

$\transph$, $\slashed{E}_T \geq 250\GeV$, $n(\ell)=0$, $p_T(j_1)\geq 250\GeV$, $p_T(j_2,j_3,j_4)\geq40\GeV$~,

\item[C5:]

$\transph$, $\slashed{E}_T\geq 150\GeV$,  $n(\ell)=0$, $p_T(j_1)\geq 150\GeV$, {$p_T(j_2, j_3,j_4)\geq40\GeV$}~,

\item[C6:]

$\transph$, $\slashed{E}_T\geq 250\GeV$, $n(\ell)=0$, $p_T(j_1)\geq 100\GeV$, $p_T(j_2)\geq 40\GeV$~,

\item[C7:]

$\transph$, $\slashed{E}_T\geq 200\GeV$,  $n(\ell)=0$, $p_T(j_1)\geq 30\GeV$~,

\item[C8:]

$\transph$, $\slashed{E}_T\geq200\GeV$, $n(j)\geq2$, $n(\ell)\geq 2$~,

\item[C9:]

$\transph$, $\slashed{E}_T\geq 200 \GeV$, $n(b\text{-jets})=1$~,

\item[C10:]

$\transph$, $\met100$, $n(\ell)=0$, $n(b\text{-jets})\geq 1$~,

\item[C11:]

$\transph$, $\met100$, $n(\ell)=0$, $n( b\text{-jets})\geq 2$~,

\item[C12:]

$\transph$, $\met100$, $n(j)\geq 4$~,

\item[C13:]

$\transph$, $\met100$, $n(j)\geq4$, $p_T(j_1)\geq 100\GeV$, $m_{\rm eff}\geq 400\GeV$~,

\item[C14:]

$\transph$, $\met100$, $n(j)\geq4$, $p_T(j_1)\geq 100\GeV$, $m_{\rm eff}\geq 550\GeV$~,

\item[C15:]

$\transph$, $\met100$, $n(j)+n(\ell)\geq4$, $p_T(j_1)\geq 100\GeV$, $H_T\geq 400\GeV$~,

\item[C16:]

$\transph$, $\met100$, $n(j)+n(\ell)\geq4$, $p_T(j_1)\geq 100\GeV$, $H_{T}\geq 550\GeV$~,

\item[C17:]

$\transph$, $\met100$, Z-veto, $n(\ell_a^+)=1$, $n(\ell_b^-)=1$, $p_T(\ell_2)\geq 20\GeV$, $p_T(j_1)\geq100\GeV$, $p_T(j_2)\geq 40\GeV$~,\footnote{
In the specification of the cuts C17 and C18, the subscripts 
$a$ and $b$ indicate that they may be different flavors, but a Z-veto is applied only to OSSF. }

\item[C18:]

$\transph$, $\met100$, Z-veto, $n(\ell_a^+)=1$, $n(\ell_b^-)=1$, $p_T(\ell_2)\geq 20\GeV$, $p_T(j_2)\geq 40\GeV$~,

\item[C19:]

$\transph$, $\met100$, $\slashed{E}_T\geq 0.2 m_{\rm eff}$, $n(j)\geq4$~, $p_T(j_1)\geq100\GeV$, 

\item[C20:]
$\slashed{E}_{T}\geq 100 \GeV$, $n(\ell)=3$, $p_T(j_1)\geq 150 \GeV$, $n(j)\geq 2$~,

\item[C21:]
$\slashed{E}_{T}\geq 150 \GeV$, $n(\ell)=3$, $p_T(j_2)\geq 40 \GeV$~.

\end{enumerate}

\begin{table}[p!]
\centering
\tiny{
 {\Large LHC significance over channels for 1~fb$^{-1}$ of integrated luminosity}
\vspace{.3cm}
\begin{center}
\begin{tabular}{|c|c|c|c|c|c|c|c|c|c|c|c|c|c|c|c|c|c|c|c|c|c|}
\hline
	&	C1&C2&C3&C4&C5&C6&C7&C8&C9&C10&C11&C12&C13&C14&C15&C16&C17&C18&C19&C20&C21	\tnhl\hline
LG1	&	6&9&18&6&19&4&5&3&7&10&8&13&17&20&16&19&2&2&17&3&2	\tnhl
LG2	&	3&4&4&13&7&14&9&10&5&2&2&4&5&8&6&8&1&1&4&9&10	\tnhl
LG3	&	4&3&1&3&2&8&7&0&2&3&1&4&3&2&2&2&0&0&1&0&0	\tnhl
LG4	&	4&2&0&1&1&5&7&0&1&0&0&1&1&1&1&1&0&0&0&0&0	\tnhl
LG5	&	4&2&0&1&1&5&6&0&1&1&0&2&1&1&1&1&0&0&0&0&0	\tnhl
LG6	&	4&3&1&3&2&8&8&0&1&1&0&3&2&2&2&2&0&0&1&0&0	\tnhl
LG7	&	6&9&17&9&23&9&9&2&12&12&11&13&17&20&16&19&0&0&17&0&0	\tnhl
LG8	&	7&10&18&9&24&9&10&2&11&13&10&15&20&21&19&21&0&1&18&1&1	\tnhl
LG9	&	0&0&1&2&1&1&1&1&1&0&0&0&1&1&1&1&0&0&1&0&0	\tnhl
LG10	&	18&24&18&13&25&29&31&4&21&23&16&30&33&28&31&27&0&0&19&1&4	\tnhl
LG11	&	12&19&31&24&39&17&15&16&24&33&42&27&34&38&32&36&2&5&34&9&5	\tnhl
LG12	&	2&4&10&16&16&8&6&3&8&6&7&6&8&11&7&10&0&1&10&2&2	\tnhl
LG13	&	6&5&8&25&14&17&13&0&11&6&5&7&7&9&7&9&0&0&7&0&0	\tnhl
LG14	&	8&10&10&24&17&31&19&25&17&6&5&11&15&20&15&20&4&5&12&20&20	\tnhl
LG15	&	9&11&16&38&26&34&22&22&19&8&7&13&18&24&18&25&5&5&16&20&18	\tnhl
LG16	&	19&28&15&11&18&11&14&0&11&27&15&37&27&16&25&15&0&0&17&0&0	\tnhl
LG17	&	7&10&13&7&19&10&11&0&11&11&8&14&17&16&16&15&0&0&14&0&0	\tnhl
 \end{tabular}
\caption{\label{discovery} 
A display of the signal significance $S/\sqrt{B}$ in each discovery channel for the models in 
Table(\ref{bench})  for 1~fb$^{-1}$ of integrated luminosity at the LHC. 
As already mentioned in  Sec.(IV) for a signal to be 
 discoverable we require $S\geq \max\left\{5\sqrt{B},10\right\}$.
}
\label{tab:1}
\end{center}
}
\end{table}

\begin{table}
\centering
\tiny{
{\Large LHC  reach for (0.5,~1,~2,~5)~fb$^{-1}$ of integrated luminosity}
\vspace{.3cm}
\begin{center}
\begin{tabular}{|c|c|c|c|c|c|c|c|c|c|c|c|c|c|c|c|c|c|c|c|c|c|}
\hline
 &       C1&C2&C3&C4&C5&C6&C7&C8&C9&C10&C11&C12&C13&C14&C15&C16&C17&C18&C19&C20&C21      \tnhl\hline
LG1	&	1.0&0.5&0.5&1.0&0.5&2.0&1.0&5.0&0.5&0.5&0.5&0.5&0.5&0.5&0.5&0.5&&&0.5&5.0&	\tnhl
LG2	&	5.0&2.0&2.0&0.5&1.0&0.5&0.5&0.5&1.0&&&2.0&1.0&0.5&1.0&0.5&&&2.0&2.0&1.0	\tnhl
LG3	&	2.0&5.0&&5.0&5.0&0.5&1.0&&&5.0&&2.0&5.0&&5.0&&&&&&	\tnhl
LG4	&	2.0&&&&&2.0&1.0&&&&&&&&&&&&&&	\tnhl
LG5	&	5.0&&&&&2.0&1.0&&&&&&&&&&&&&&	\tnhl
LG6	&	2.0&5.0&&5.0&&0.5&0.5&&&&&5.0&&&&&&&&&	\tnhl
LG7	&	1.0&0.5&0.5&0.5&0.5&0.5&0.5&&0.5&0.5&0.5&0.5&0.5&0.5&0.5&0.5&&&0.5&&	\tnhl
LG8	&	1.0&0.5&0.5&0.5&0.5&0.5&0.5&&0.5&0.5&0.5&0.5&0.5&0.5&0.5&0.5&&&0.5&&	\tnhl
LG9	&	&&&5.0&&&&&&&&&&&&&&&&&	\tnhl
LG10	&	0.5&0.5&0.5&0.5&0.5&0.5&0.5&2.0&0.5&0.5&0.5&0.5&0.5&0.5&0.5&0.5&&&0.5&&5.0	\tnhl
LG11	&	0.5&0.5&0.5&0.5&0.5&0.5&0.5&0.5&0.5&0.5&0.5&0.5&0.5&0.5&0.5&0.5&5.0&2.0&0.5&2.0&2.0	\tnhl
LG12	&	5.0&2.0&0.5&0.5&0.5&0.5&1.0&5.0&0.5&1.0&1.0&1.0&0.5&0.5&0.5&0.5&&&0.5&&	\tnhl
LG13	&	1.0&1.0&0.5&0.5&0.5&0.5&0.5&&0.5&1.0&2.0&1.0&0.5&0.5&1.0&0.5&&&0.5&&	\tnhl
LG14	&	0.5&0.5&0.5&0.5&0.5&0.5&0.5&0.5&0.5&1.0&2.0&0.5&0.5&0.5&0.5&0.5&2.0&1.0&0.5&0.5&0.5	\tnhl
LG15	&	0.5&0.5&0.5&0.5&0.5&0.5&0.5&0.5&0.5&0.5&1.0&0.5&0.5&0.5&0.5&0.5&2.0&1.0&0.5&0.5&1.0	\tnhl
LG16	&	0.5&0.5&0.5&0.5&0.5&0.5&0.5&&0.5&0.5&0.5&0.5&0.5&0.5&0.5&0.5&&&0.5&&	\tnhl
LG17	&	1.0&0.5&0.5&0.5&0.5&0.5&0.5&&0.5&0.5&0.5&0.5&0.5&0.5&0.5&0.5&&&0.5&&	\tnhl
\end{tabular} 
\caption{
The values in the table are the integrated 
 luminosity in units of fb$^{-1}$ 
when the model is first discoverable in that channel.
The table shows that many of the low mass gluino models 
will become visible with an integrated luminosity of 0.5~fb$^{-1}$,
and all models become visible with an integrated luminosity of 2~fb$^{-1}$ 
except the model LG9 which requires
an integrated luminosity of 5~fb$^{-1}$ to be discovered under the  criterion given in
Sec.(IV) and in the caption of Table(IX).
\label{grid}}
\label{tab:all}
\end{center}
}
\end{table}

\clearpage

{\it 3. Signature Analysis: }   Our analysis is carried out to determine the potential for  discovery  of the dark matter motivated models $\rm LG1,\ldots,LG17$,
for  0.5~fb$^{-1}$, 1~fb$^{-1}$, 2~fb$^{-1}$ 
and  5~fb$^{-1}$ (with a focus on 1~fb$^{-1}$) of integrated luminosity with 7 TeV center of mass energy.  These models exhibit generic features of a very broad class of SUSY models. 

 We now discuss  Fig.(\ref{fig:significance}), shown in Sec.(I), a bit more generally. The potential for discovering the models of Table (\ref{bench})  at  1~fb$^{-1}$ of integrated luminosity 
 is given in Table(\ref{tab:1}). A subset of the results in Table(\ref{tab:1}) are given in  Fig.(\ref{fig:significance}).
Thus, Table(\ref{tab:1})   
 displays each model's significance, $S/\sqrt{B}$, for the 21 cuts listed above. One finds that a particular model, ${\rm LG}_{k}$, often has several signatures that lead to large  excesses of signal over background,  i.e. $S/\sqrt B >5$, and the set of signatures in which 
the model becomes visible varies significantly from one model to the next. 
Additionally, in studying the tri-leptonic channels, C20 and C21, we see that consistently only 6 models  among those listed in Table(\ref{bench}) are discoverable (LG1, LG2, LG10, LG11, LG14, LG15) and the majority of the other models show less then 10 events in the tri-leptonic channels. 

\begin{figure*}[t!]
\begin{center}
\includegraphics[scale=0.35]{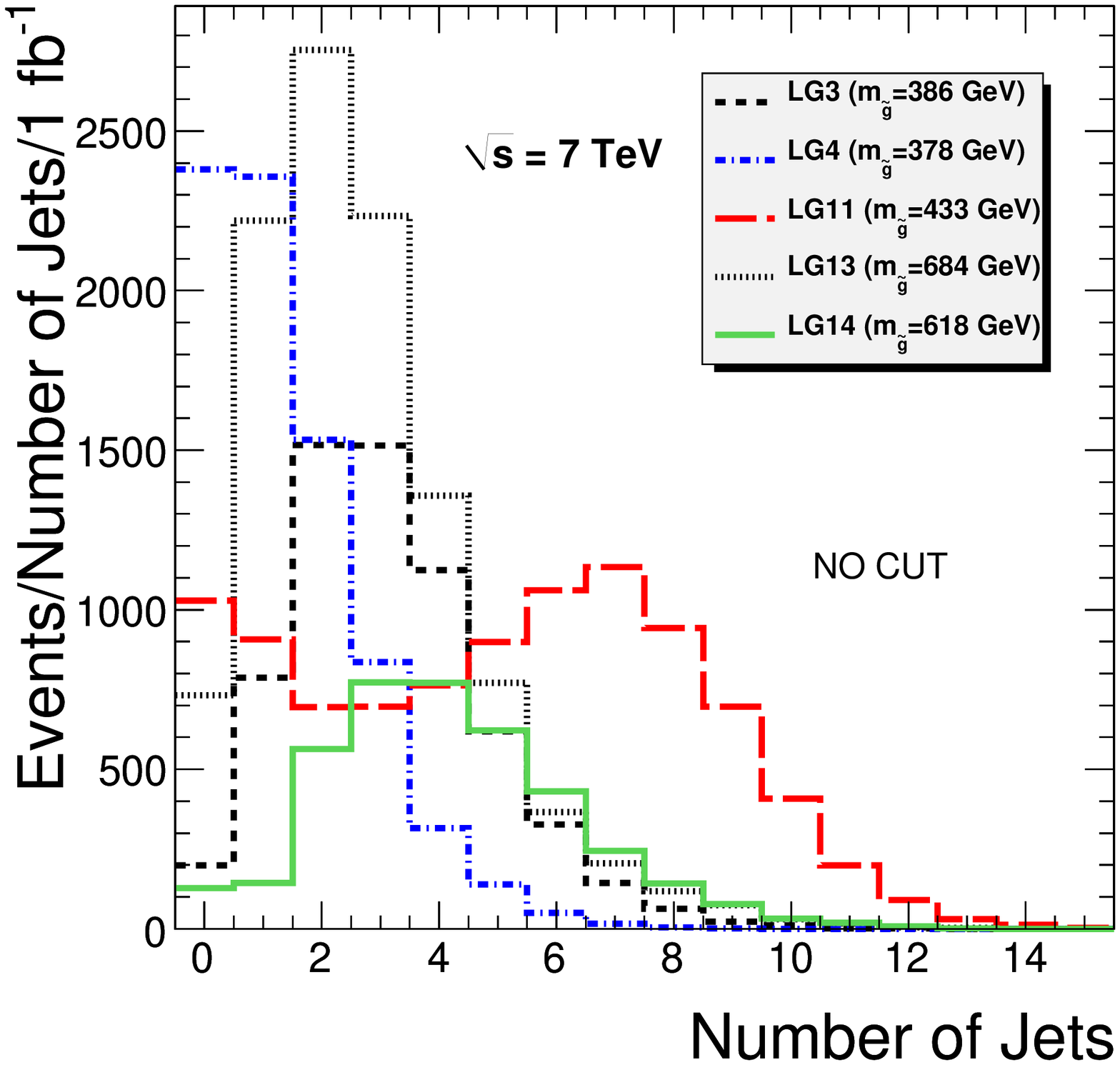}
\includegraphics[scale=0.35]{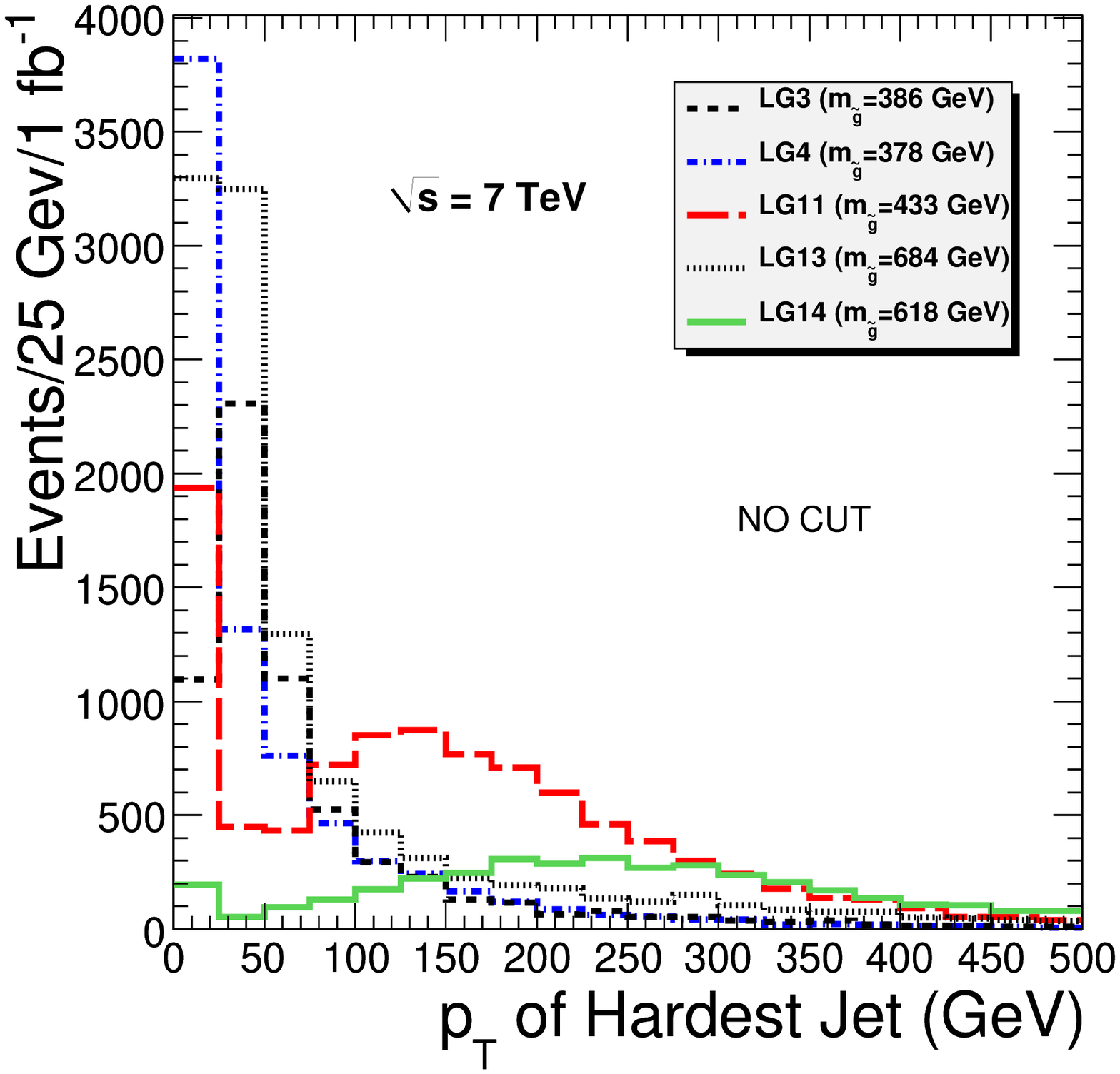}
\includegraphics[scale=0.35]{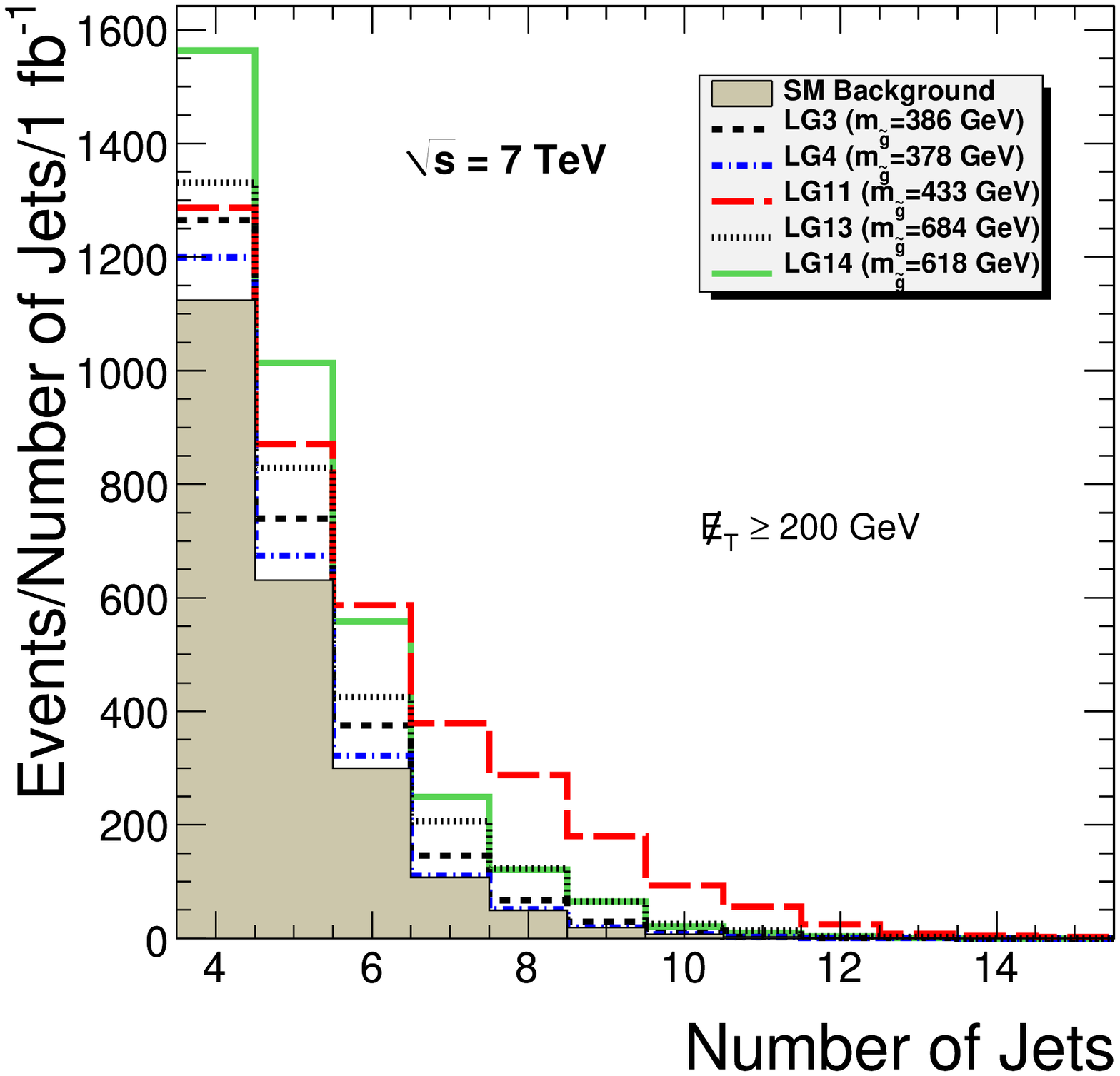}
\includegraphics[scale=0.35]{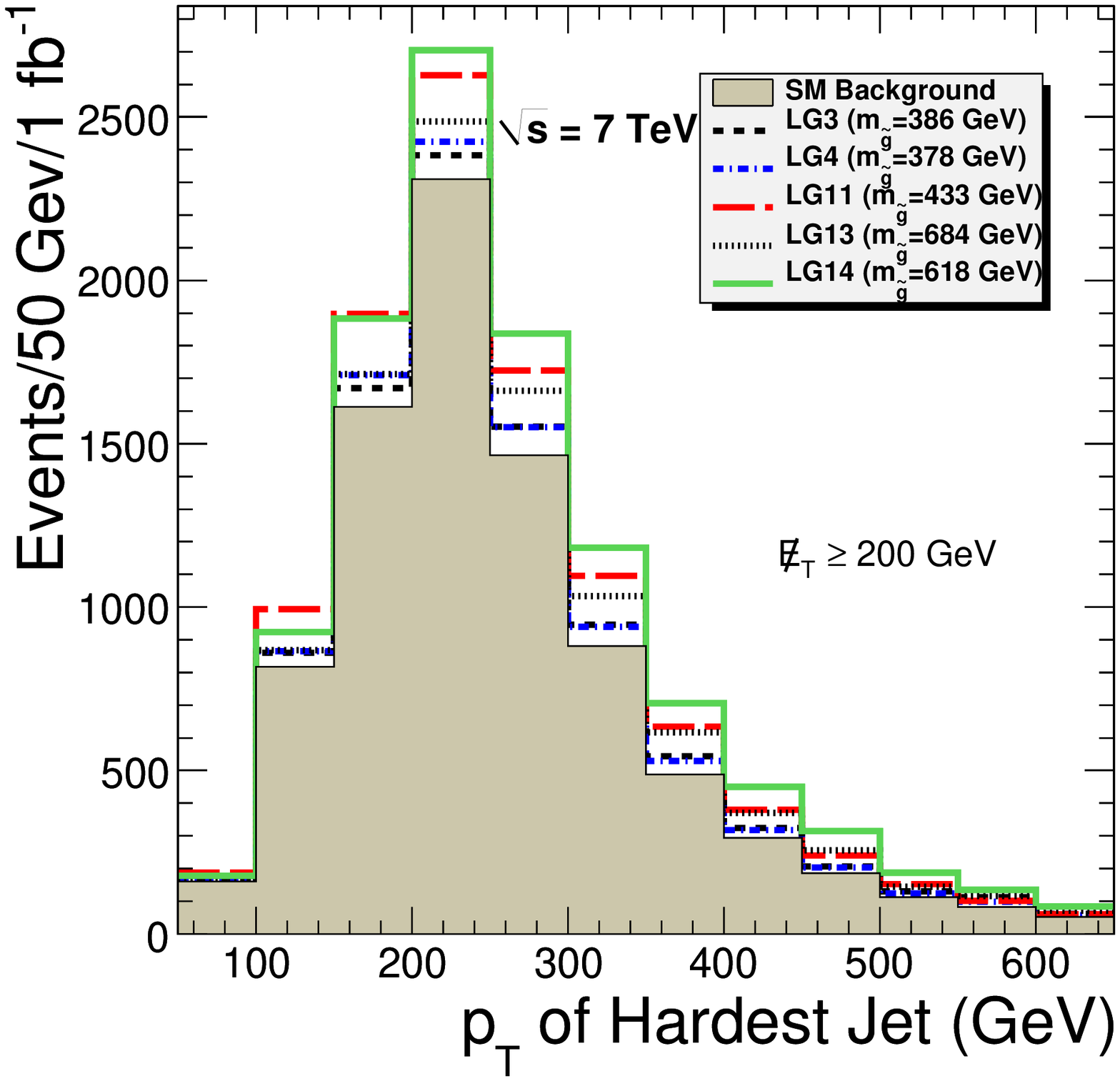}
\caption{(Color online) Top Left: Distribution of the number of jets 
without cuts. Top Right: Distribution of the $p_T$ of the 
hardest jet also without cuts.  Bottom Left: Distribution of the number of SUSY events (plus SM background) vs.   
the number of jets after a cut of $\slashed{E}_T\geq 200\GeV$.  Bottom Right: Distribution of the number of SUSY events (plus SM background) vs. the  $p_T$ of the hardest jet after a cut of $\slashed{E}_T\geq 200\GeV$. 
}
\label{fig:jet}
\end{center}
\end{figure*}

The overall production cross section of superparticles is determined mainly
by the production of squarks and gluinos. For the gluino production modes at the LHC, 
the cross 
section is determined by the gluino mass. However, for   models with low mass gluinos, like those that appear in Table(\ref{bench}), the detectable signals at the LHC are strongly influenced by the other low-lying 
superparticles, i.e., the superparticles that are lighter than the gluino. 
In Fig.(\ref{fig:jet}), an analysis of the jet multiplicity and the transverse momentum 
of the leading jets is given for LG3, LG4, LG11, LG13, and LG14. The distributions for jet multiplicity and jet momentum look 
quite different from model to model.  For instance, the model LG11 has a gluino mass of 433 GeV, and several of its superparticles are lighter than the gluino, including 
the lighter stop and gauginos.  This leads to lengthy cascade decay chains which produce multiple jets.
Further, the mass differences between the superparticles in LG11 are 
relatively large which give rise to large momentum of the SM final states including jets. 
In contrast,  models LG3 and LG4 tend to produce events with less jet multiplicity 
and smaller transverse momentum, which is due to the fact that  these  models having a gluino as the NLSP; i.e., these models are 
 GNLSPs. For these GNLSP models the masses of the gluino and the LSP are 
correlated in the gluino coannihilation mechanism  such that 
the mass gap is relatively small.

Specifically,  LG3 has $M_{\tilde g}-M_{\na} \sim 50 ~\rm GeV$
and the gluino production is, overwhelmingly, the dominant production
mode. For this model, the gluinos decay
directly to the LSP + 2 jets, i.e., $ {\mathcal Br}(\g \to (b \bar b
\na, q\bar{q} \na ))\sim (20,80) \%$ where $q$ stands for 
first two generation quarks. We note in passing that generally one needs
to take into account the radiative decay of the gluino, $\g \to g \na$. 
 Such a case occurs, for example, in LG4, and the decay
 $\g \to g \na$  dominates the branching ratio. 
 The relatively small mass splitting in model LG3 between the gluino and the LSP (as well as the
extreme case of LG4) makes this model class harder to discover
 due to the softer jets and low jet multiplicity, compared to other models. 
 (For recent work on relatively small gluino-LSP splittings see \cite{land2,land3,GNLSP} and \cite{Wacker,Wells}.)  This feature is illustrated further in  Table(\ref{discovery}) and Table(\ref{grid}).
We note that in such cases where the mass gap between gluino and the LSP is  
extremely small, the effects of the ISR can be substantial for the collider signatures. 
We also note that although it can be challenging to discover events from gluino production for the  GNLSP models,
 (depending on the degree of the mass degeneracy),
one should keep in mind that some other subdominant SUSY production modes could 
be detectable and become the leading signals for such models. For example, 
in model LG3, the stop is relatively light and decays entirely into a chargino and bottom quark, i.e.,
 ${\mathcal Br}(\ta \to  \tilde{\chi}^{+}_1 b )\sim 100\%$,
and the chargino subsequently decays entirely into a neutralino and a $W$ boson, i.e.,  $ {\mathcal Br}( {\tilde{\chi}^{+}_1} \to \na W^+) \sim 100 \%$. 
Hence, this GNLSP model class may be able to produce
discoverable leptonic events through these decay chains
with upgraded center of mass energy and luminosity.

 Further, these features of the GNLSP models can explicitly be seen by studying the top panels of Fig.(\ref{fig:jet})
and by observing the relative broadness (or width) of the  $p_{T}$ distribution  
of models LG11 and LG14 relative to LG3 and LG4.
 In addition, LG13, a stop NLSP  model, is also peaked at low jet $p_T$ much like LG3 and LG4.  The stop mass for model LG13 is near 200~GeV and the stop-LSP mass splitting is small ($\lesssim 30$~GeV).  Thus, this model produces stops at a large rate, which decay (via an off-shell loop-induced and FCNC decay) into a charm quark and LSP $(\ta \to c\na)$ with a  $\sim 75\%$  branching ratio.
However, the softness of  jets in LG13 mimics the softness of jets in LG4
 in part due to the phase space, which
explains the peaking of the distributions at low $p_T$.  The restricted phase space from the small mass splittings
is also why  the effective mass distribution for stop NLSPs
 is narrow (see \cite{land3}) relative to other cases.

 \begin{figure*}[t!]
 \begin{center}
    \includegraphics[scale=0.35]{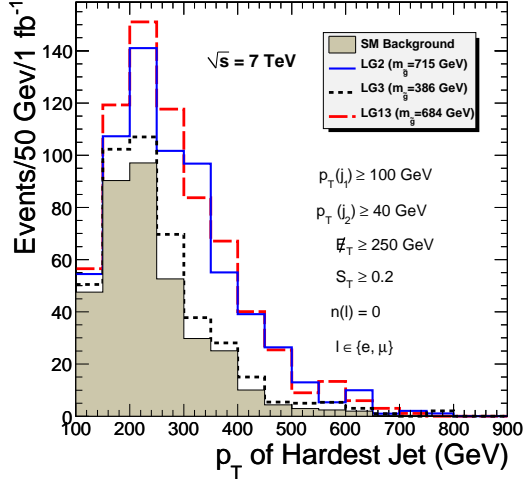}
\caption{
SUSY plus SM background events vs $p_{T} (j_1)$ at 1~fb$^{-1}$ for the signature cut $p_T\left(j_1\right)\geq 100\GeV$, $p_T\left(j_2\right)\geq 40\GeV$, $\slashed{E}_T\geq 250 \GeV$, $S_{T}\geq0.2$ and $n(\ell)=0$ for 
 LG2, LG3, LG13. The figure illustrates the softness of the jets in model LG3, a GNLSP model, relative to the models LG2 and LG13.  }
\label{ptjetLG3}
\end{center}
\end{figure*}

In Fig.(\ref{ptjetLG3}) we highlight the GNLSP model LG3, which satisfies the double-sided relic density band, along with the light stau and stop models,
LG2 and LG13, which also satisfy the WMAP bound via scalar coannihilations.
Thus, Fig.(\ref{ptjetLG3}) shows jet  $p_T$, signal  plus  background, for the models LG2, LG3, and LG13 compared to the SM background alone.
  As discussed earlier the model LG3  arises from gluino coannihilations
 and has a relatively small mass splitting between the gluino and the LSP neutralino. This is to
 be contrasted with  the model LG2, which satisfies the WMAP relic density band via stau coannihilations, or  the model LG13, which also
 satisfies the WMAP relic density band via  stop 
 coannihilations. Because of the compressed spectra of LG2 and LG13, there are  more jets arising from the
 combinations of both low mass gluino and the low mass squark production relative to the dominant gluino production
 found in the GNLSP model LG3. This effect is exhibited in the figure. For model LG2,  as the scalars are quite light, and even lighter than the
715 GeV gluino, the   cross section for  the production of squarks as
well as the  mixed squark gluino production cross sections
are about an order of magnitude larger than the $\g \g$ production.
Here the gluino two body decay modes are spread out rather uniformly
with no dominant channel. Instead the first two generation
squark decay modes are short with large branchings. In particular,
one has  for the first two generation squarks, ${\mathcal Br} (\tilde{q}_{R} \to
\na q) \sim 100 \% $
and ${\mathcal Br}(\ql \to \nb q) \sim 32 \% $ as well as
 ${\mathcal Br} ((\tilde{q}_{d_L}, \tilde{q}_{u_L}) \to{\tilde{\chi}^{(-,+)}_1} (q_u,q_d)) \sim
(60-65) \% $ for
 each decay. Thus, the two body decays of the first two generation
squarks provide
 the large signal in model LG2 even though the gluino is quite light.
 In addition, for LG2, the direct production of chargino pairs as well as chargino
and neutralino
 production is competitive with the squark production, leading to
leptonic decays
 and large lepton multiplicities.  
The discovery potential of the models is also exhibited  
in Fig.(\ref{fig:significance}), where one finds that the significance of LG3 and LG4 is
much less than for LG11.

  \begin{figure*}[t!]
 \begin{center}
    \includegraphics[scale=0.35]{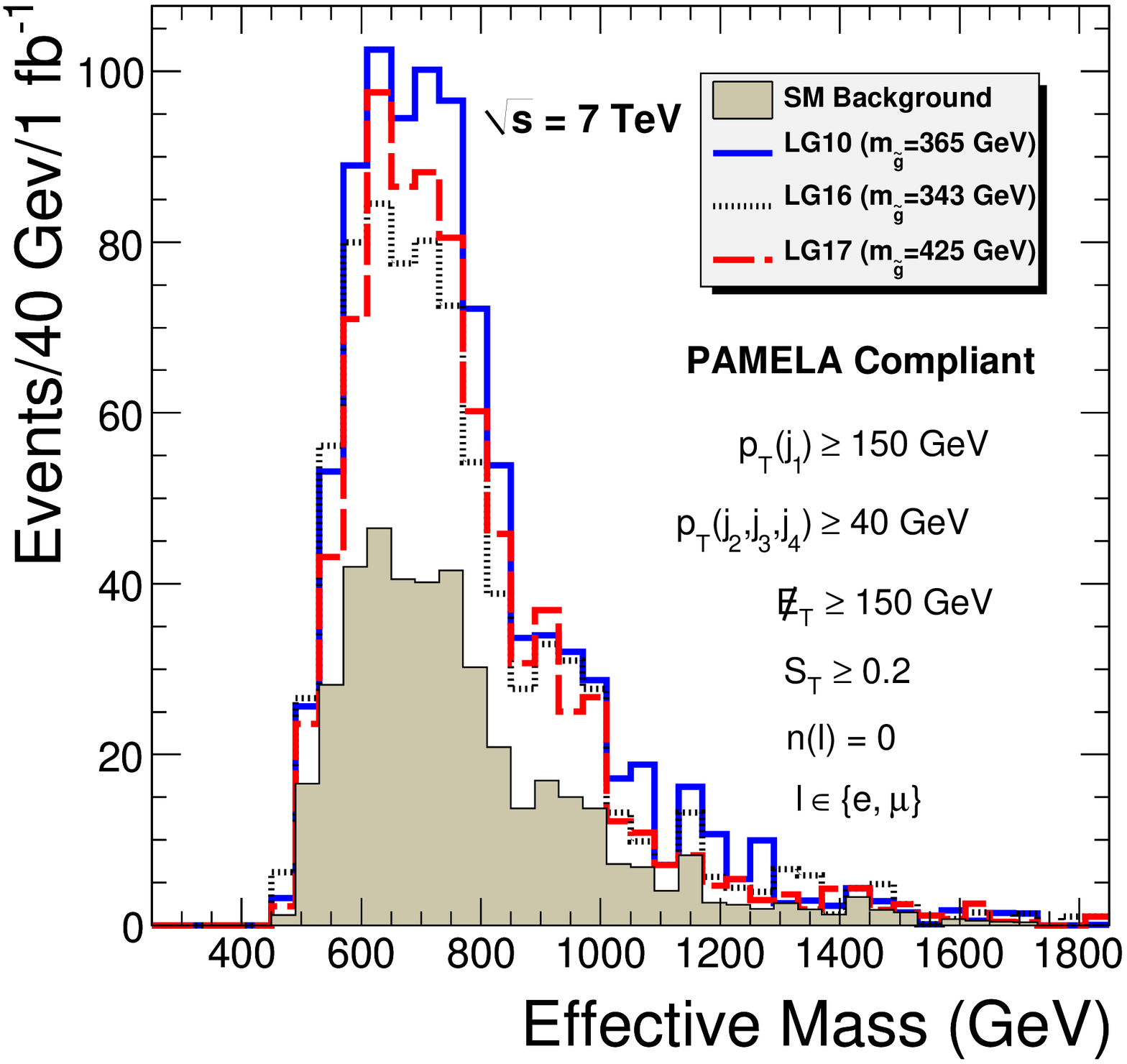}
   \includegraphics[scale=0.35]{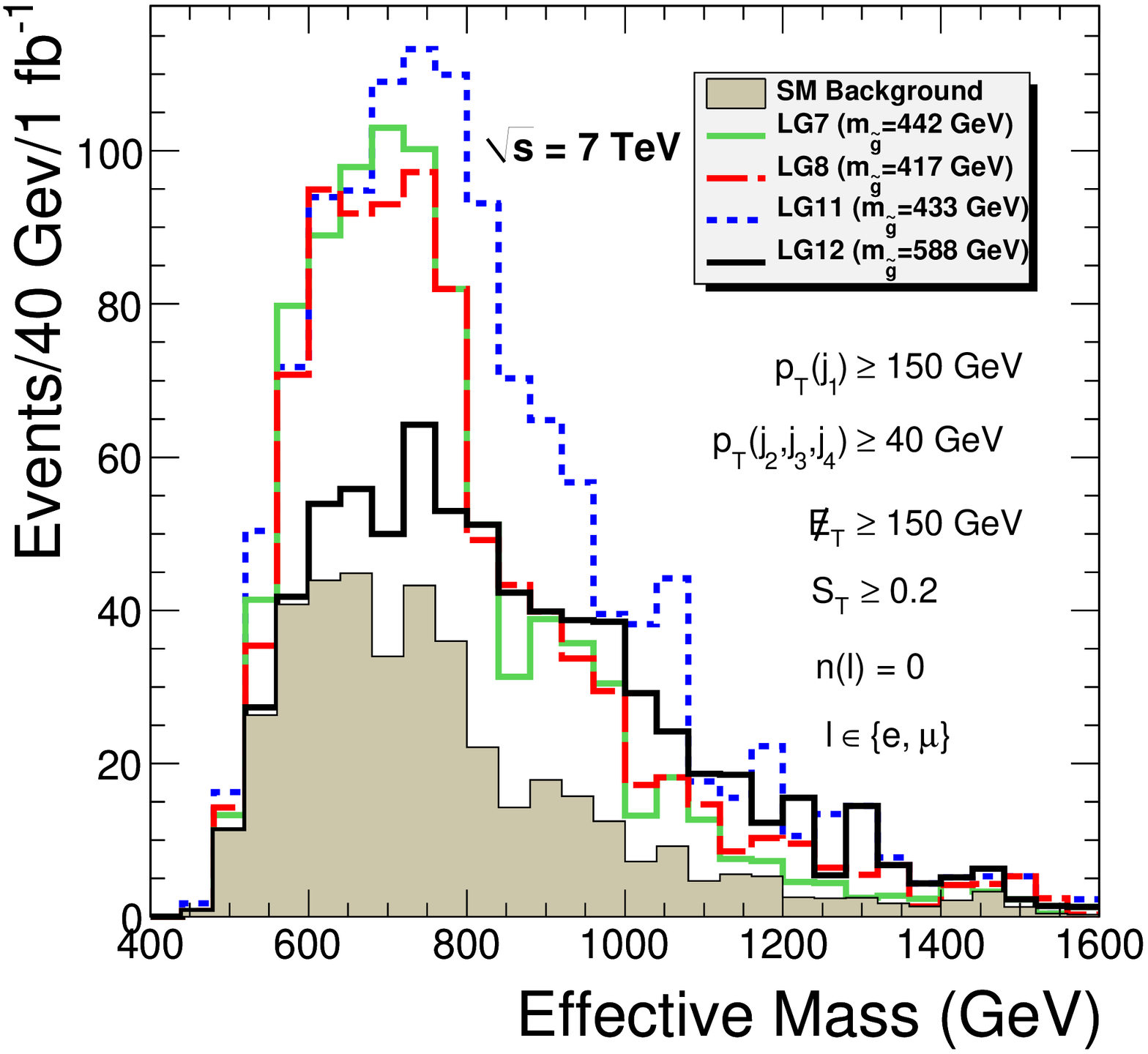}
\caption{
Left: SUSY plus SM background events vs $m_{\rm eff}$ at 1~fb$^{-1}$ 
of integrated luminosity 
for the signature cut $p_T\left(j_1\right)\geq 150\GeV$, $p_T\left(j_2,j_3,j_4\right)\geq 40\GeV$, $\slashed{E}_T\geq 150 \GeV$, $S_{T}\geq0.2$ and $n(\ell)=0$ for 
 the PAMELA compliant models.  As discussed in the text  LG10 is a Higgsino LSP model and  LG16 and LG17 are models with a mixed-wino LSP. Right: The same as the left panel except for  a subset of the  GNNLSP models (with chargino
 and neutralino degenerate),
 i.e., LG7, LG8, along with the compressed models LG11, LG12, which in addition to a low mass gluino, also have  a light stau and a light
  stop and have a compressed mass spectrum for the 
   first two generation squarks and sleptons. }
\label{pamlhc}
\end{center}
\end{figure*}

In Fig.(\ref{pamlhc})  (left-panel) we show the potential for 
early discovery  for the PAMELA compliant  model class discussed in Table(\ref{pamela})
  at  $\sqrt{s}=7$ TeV and at an  
  integrated luminosity of $1~\text{fb}^{-1}$. The displayed models have a  rather
large gluino production.
  Since the gluino, the light chargino, and the 
second heaviest neutralino   are the 
lightest SUSY particles beyond the LSP, and the squarks are rather heavy for these models,
the sparticle production at the LHC will be  dominated by 
$\g \g$,  
 $\nb \cha$  and $\chi^{+}_1\chi^{-}_1$ production.
   For example,   models (LG10, LG16, and LG17) have a total SUSY 
cross section of $\sim (12,15,5)$~pb  at  leading order  and the
 gluino production is at the level of $\sim (9,14,4)$~pb, respectively. 
  The chargino neutralino production makes up most of the remaining part of the cross section. 
 The  leading decays of the gluino are 
 $ \g \to \tilde{\chi}_{1}^{\pm} +\bar{q}q'$  and 
 $\g \to \nb/\na + q  \bar q$.
 These decays are subsequently
followed by $\nb \to \na  +\bar{f}f$ and 
$\tilde{\chi}_{1}^{\pm} \to \na + \bar{f}f'$  where $f, f'$ are the standard model quarks and leptons.
  In particular, the lightness of the gluino in the three models (LG10, LG16,and LG17) gives rise to multi-jets
which produce a strong signal over the background.  Hence, these models 
are good candidates for early discovery.
Further, if a model of the type  LG10, LG16, or LG17 is verified at the LHC, it would also provide 
a consistent explanation of the  PAMELA anomaly.
 However, to fully demonstrate the validity of the models, additional luminosity would be needed to extract 
information about the neutralino mass. We do not give a detailed methodology for accomplishing this, but as 
 argued in  \cite{Feldman:2010uv} it may be possible to extract information about the neutralino and 
 the chargino states in the gluino decay products.

\begin{figure}[t!]
\begin{center}
\vspace{.3cm}
\includegraphics[scale=0.35]{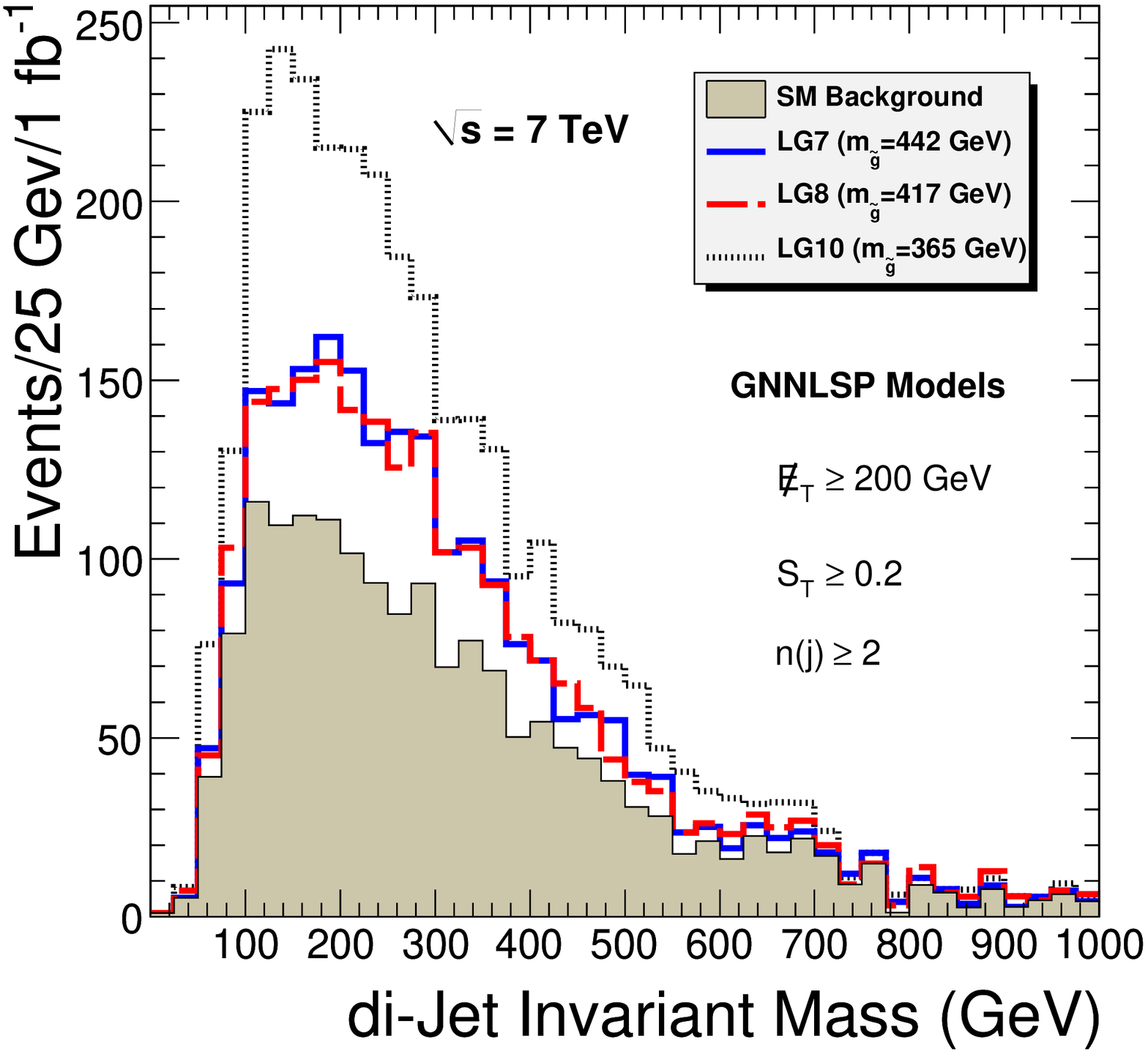}
\includegraphics[scale=0.35]{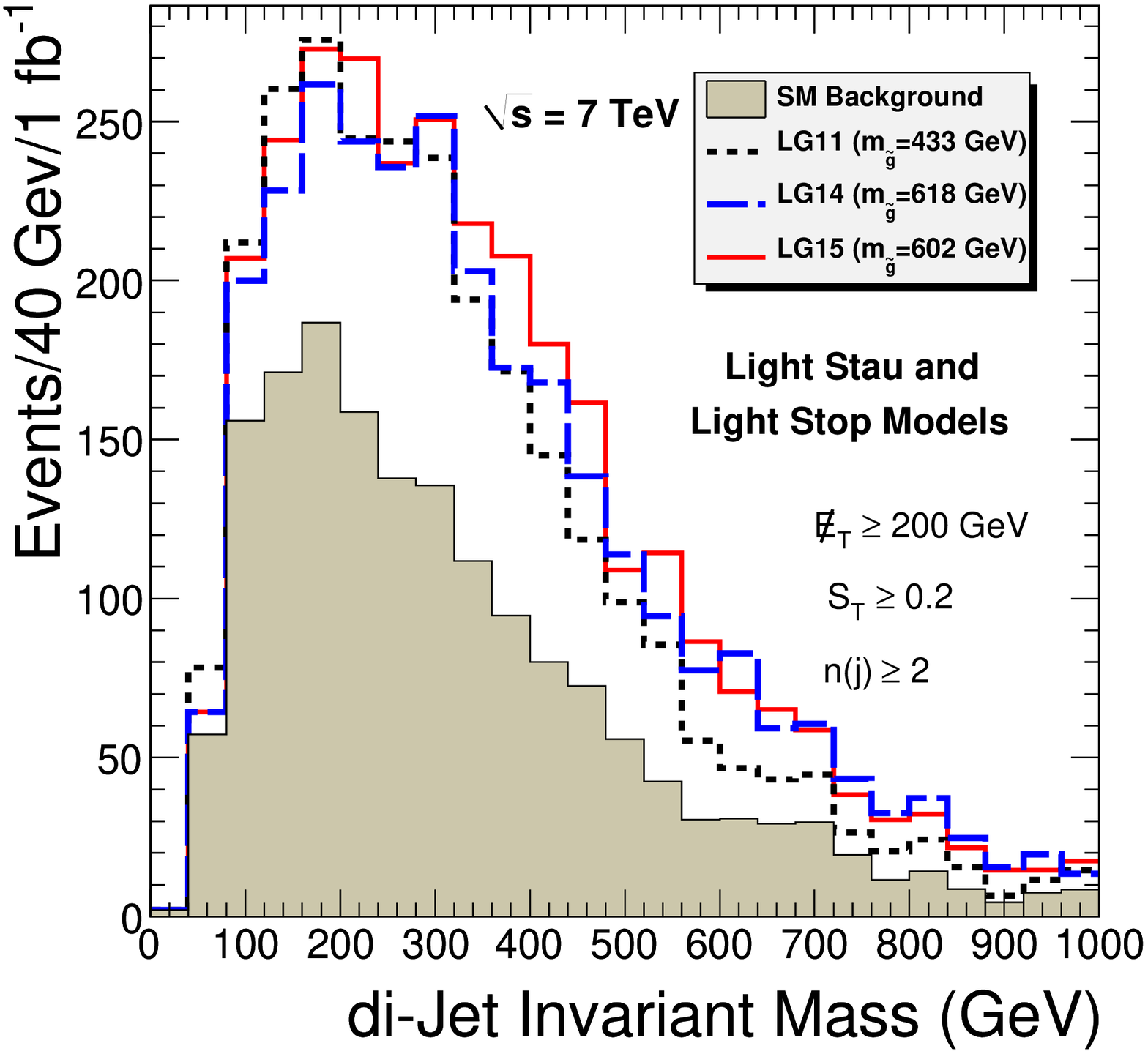}
\caption{
Left: SUSY plus SM background events  vs the di-jet invariant mass ($m_{jj}$) at 1~fb$^{-1}$
of integrated luminosity
 for signature cut 
 $\slashed{E}_T \geq 200 \GeV$, $S_T\geq 0.2$ and $n(j)\geq 2$ for  the models LG7, LG8, LG10.
 Right: Same as the left plot except that the analysis is for models 
  LG11, LG14, LG15. The left panel shows the light gluino models
 which are effectively GNNLSP models, while the right panel shows the models with
 a compressed mass spectrum for the  scalars and for the light gluinos.
   As such the
 right panel shows distributions which are significantly broader from the squark production and decays.   }
\label{hist3}
\end{center}
\end{figure}

In  Fig.(\ref{hist3}), we show a comparison of di-jet invariant mass distributions for the GNNLSP models compared to models where the gluino is positioned higher in the mass hierarchy.  One sees the GNNLSP models (LG7, LG8, and LG10) have a relatively narrow di-jet invariant mass which arises from these models being dominated by the three-body decays resulting from $\nb\cha$, $\tilde{\chi}_1^+\tilde{\chi}_1^-$, and $\tilde{g}\tilde{g}$ production.  Further, the distributions for the models LG8 and LG10 become depleted (or more narrow) relative to the light stau and light stop models  from the three-body decays which result in softer jets.  These subsequent decays produce an increase in the multi-jet signal compared to the SM background.  However, the light stau and light stop models LG11, LG14, and LG15 have a relatively broader distribution, which arises from the compression of their sparticle spectrum.  For these models, all the sparticle masses are below 700~GeV.  Further, LG14 and LG15, where the mass spectra are compressed, 
 the gluino is in the $31^{\rm st}$ position of the mass hierarchy.  The compressed spectra causes a large sampling of sparticle production, which results in a production of many jets with a more diverse range of momentum.

 \begin{figure}[t!]
\begin{center}
\vspace{.3cm}
\includegraphics[scale=0.35]{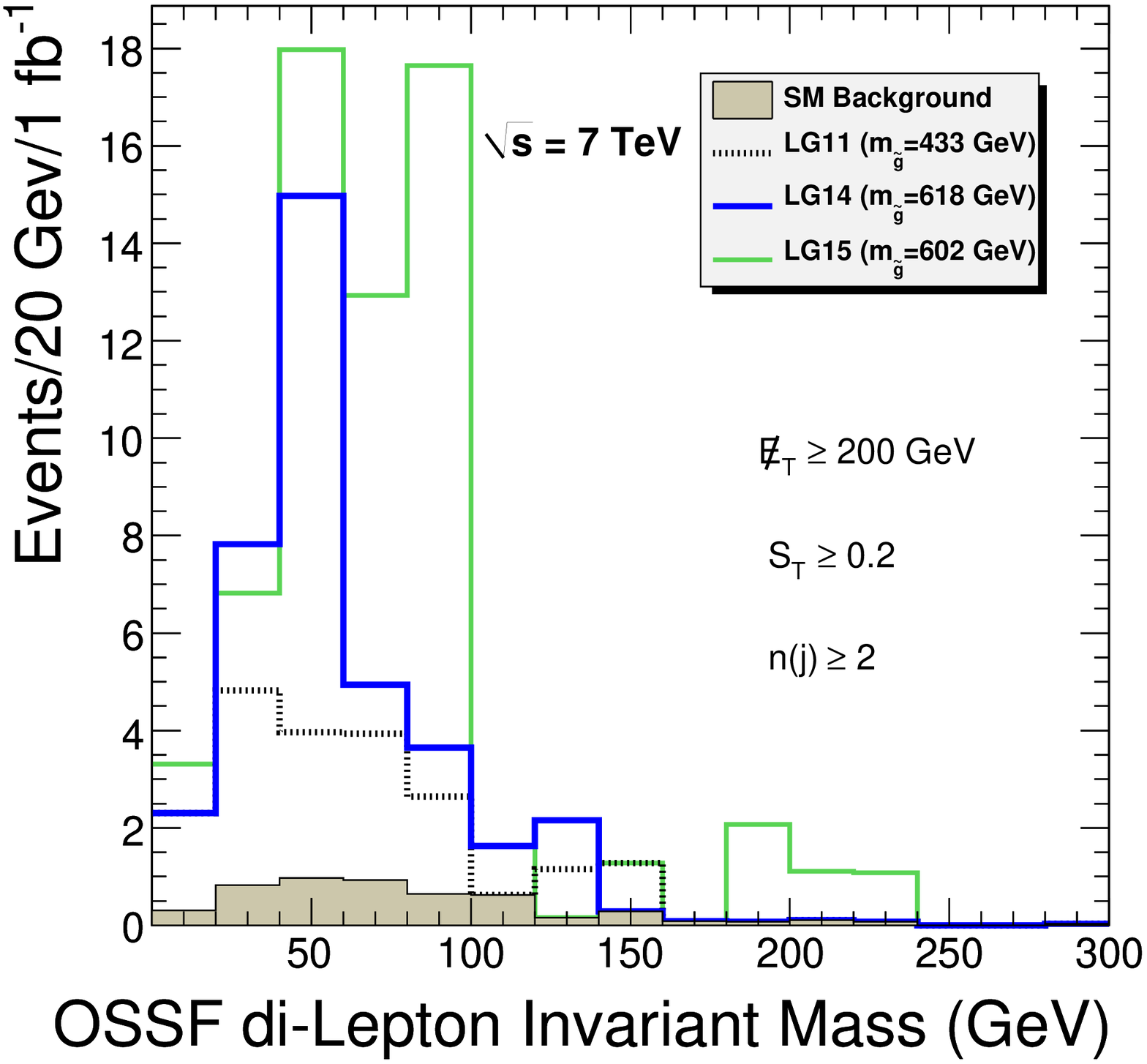}
\includegraphics[scale=0.35]{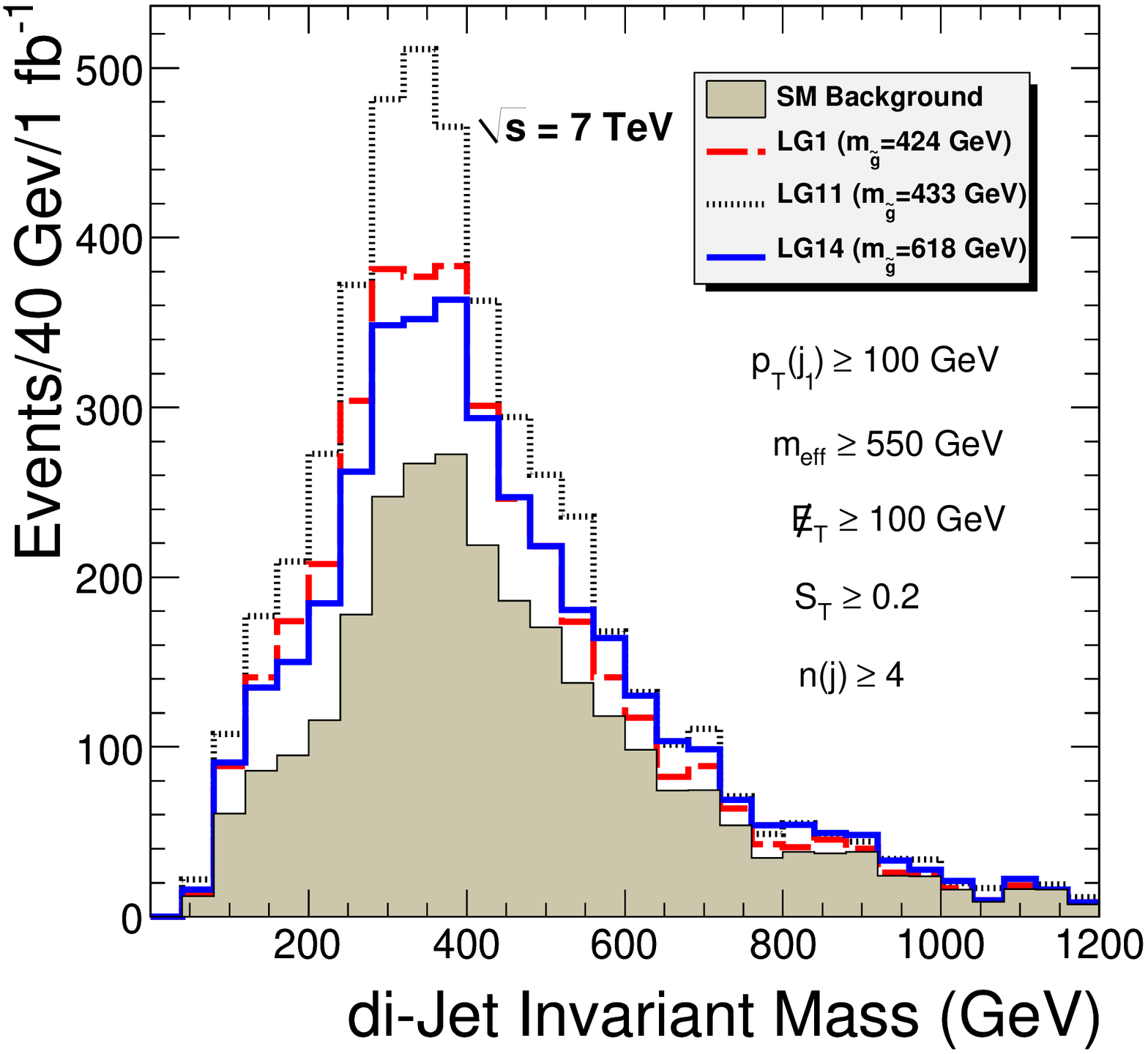}
\caption{
Left: SUSY plus background events for models LG11, LG14, LG15 
 vs the OSSF di-lepton invariant mass ($m_{\ell^+ \ell^-}$) at 1~fb$^{-1}$ for signature 
 cut 
 $\slashed{E}_T \geq 200 \GeV$, $S_T\geq 0.2$ and $n(j)\geq 2$ 
 with 2 leptons of any sign and flavor.
  Right:  SUSY plus background events  for models LG1, LG11, LG14
 vs the di-jet invariant mass ($m_{jj}$) at 1~fb$^{-1}$ of integrated luminosity
  for signature cut $\slashed{E}_T\geq 100 \GeV$, $S_{T}\geq 0.2$, $p_{T}(j_1)\geq 100 \GeV$, $m_{\rm eff}\geq 550 \GeV$ and $n(j)\geq 4$.  Here the peak in the distribution is a consequence of the $m_{\rm eff}$ cut.}
 \label{hist1}
\end{center}
\end{figure}

The right panel of Fig.(\ref{hist1})  shows the number of
SUSY signature events plus the SM background in 40~GeV energy bins at  1~fb$^{-1}$ of integrated luminosity vs  the  di-jet
 invariant mass for models LG1, LG11, and LG14.  As exhibited in this figure, these models have a significantly larger di-jet invariant mass compared to the SM.  As discussed earlier LG11 has a relatively large sparticle mass splittings in the scalar sector relative to the LSP mass as well as lengthy cascade decay chains that produce multiple final state jets with large momentum.  Further, the right panel of Fig.(\ref{hist1}) helps illustrate the effectiveness of the $m_{\rm eff}$ cut.  Comparing the values of C13 to C14 in Table(\ref{discovery}), one sees that the significance for models LG1, LG11, and LG14 increases as $m_{\rm eff}$ increases.  For the case when $m_{\rm eff} \geq $  400~GeV (C13) we get $S/\sqrt{B}=(17,34,15)$ and when $m_{\rm eff} \geq$  550~GeV (C14) we get $S/\sqrt{B}=(20, 38, 20)$, respectively, for (LG1 ,LG11, LG14).  However, models LG3, LG10, and LG16 have a reverse effect, i.e., $S/\sqrt{B}=(3,23,27)$ for $m_{\rm eff} \geq$  400~GeV (C13) and $S/\sqrt{B}=(2,15,15)$ for $m_{\rm eff} \geq $  550~GeV (C14) for (LG3, LG10, LG16), respectively.  These effects arise since models LG3, LG10, and LG16 have lower jet multiplicity, less missing energy, and fewer cascades than the models shown in the right panel of Fig.(\ref{hist1}). For instance, the model LG16 cross section is dominated by $\tilde{g}\tilde{g}$ production with the $\tilde{g}$ dominantly decaying into $\na$ or $\cha$.  This results in low jet multiplicity and lower missing energy compared to models LG1, LG11, and LG14.

{\it 4. Mass Reconstruction:}
 We now discuss the potential to do mass reconstruction for some of the models with the early data.  In the left panel of Fig.(\ref{hist1}) we display the number of
SUSY signature events plus background events  in 20 GeV energy bins at  1~fb$^{-1}$ of integrated luminosity vs  the OSSF di-lepton invariant mass for models LG11, LG14, and LG15. The plot also displays the cuts used as well as the standard model background alone for comparison. In large portions of the figure, the SUSY signals
plus the background distribution stands significantly above the  background.  The leptonic events are mostly produced from the gaugino cascade decays through low-lying sleptons.
If the OSSF di-leptons arise from the same decay chain 
$\nb\rightarrow~\tilde{\ell}^{\pm}\ell^{\mp} \rightarrow~\na\ell^{\pm}\ell^{\mp}$, 
the invariant mass from the reconstruction of the di-leptons obeys the following 
mass relations for on-shell sleptons: 
\beqn\label{ossf_edge}
M_{\ell^{+}\ell^{-}}\le M_{\nb}\sqrt{1-\frac{M_{\tilde{\ell}}^{2}}{M_{\nb}^{2}}}\sqrt{1-\frac{M_{\na}^{2}}{M_{\tilde{\ell}}^{2}}}~.
\eeqn
In particular, for model LG15 the three sleptons, $(\sta,\ser,\smr)$ with the latter two being degenerate,  contribute to the 
OSSF di-lepton events with the di-lepton invariant mass lying between $\na$ and $\nb$. 
Using the sparticle masses from LG15, i.e., 
$(M_{\na},M_{\tilde{\ell}_R},M_{\nb})=(121,137,240)$ GeV, 
one can use Eq.(\ref{ossf_edge}) to obtain   $M_{\ell^{+}\ell^{-}}\lesssim 92 \GeV$ 
from the $\ser/\smr$ decay modes.
The $\nb$ decays into $\na$ via   $\mathcal{B}r(\nb\rightarrow \tilde{\ell}_{R}^{\pm}\ell^{\mp})
\simeq 28\%$,  and then the right-handed slepton decays entirely into a lepton and $\na$, i.e.,
$\mathcal{B}r(\tilde{\ell}_R^{\pm}\rightarrow\na \ell^{\pm})
\simeq 100\%$.  The decay of the light stau follows similarly through
 $\mathcal{B}r(\nb\rightarrow\sta^{\pm}\tau^{\mp})\simeq 33\%$, and then the stau decays entirely into $\na$ and a $\tau$, i.e.,  $\mathcal{B}r(\sta^{\pm}\rightarrow \na \tau^{\pm})
\simeq 100\%$.  The  tau produced from the $\nb$ decay has a subsequent leptonic tau decay with branching ratio of 
$\mathcal{B}r(\tau\rightarrow \ell {\nu}_{\ell}\nu_{\tau})
\simeq 35\%$. 
We do not attempt to reconstruct taus here, and we note that there are also further mixings arising from chargino decays which require flavor subtraction and other 
techniques to isolate lepton pairs coming from the same cascade decay.  Further, due to the low statistics 
at the early runs of the LHC data, we do not perform a more detailed 
mass reconstruction in our analysis here. As discussed above,  and as can be seen in Fig.(\ref{hist1}), 
the mixings arising from other processes are rather small and the edge in the di-lepton 
invariant mass agrees well with the prediction of Eq.(\ref{ossf_edge}).

\begin{figure}[t!]
\begin{center}
\vspace{.3cm}
\includegraphics[scale=0.35]{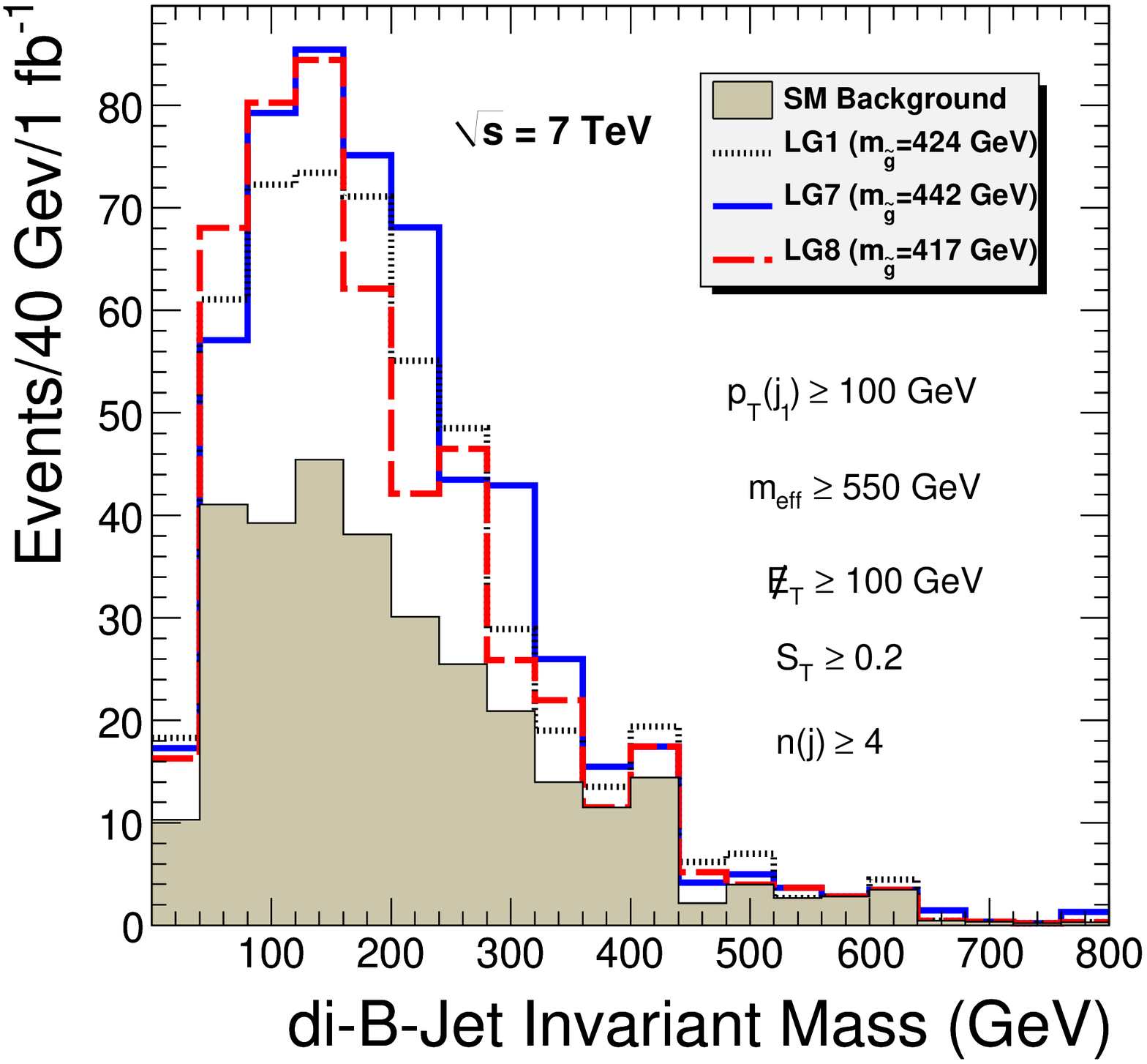}
\includegraphics[scale=0.35]{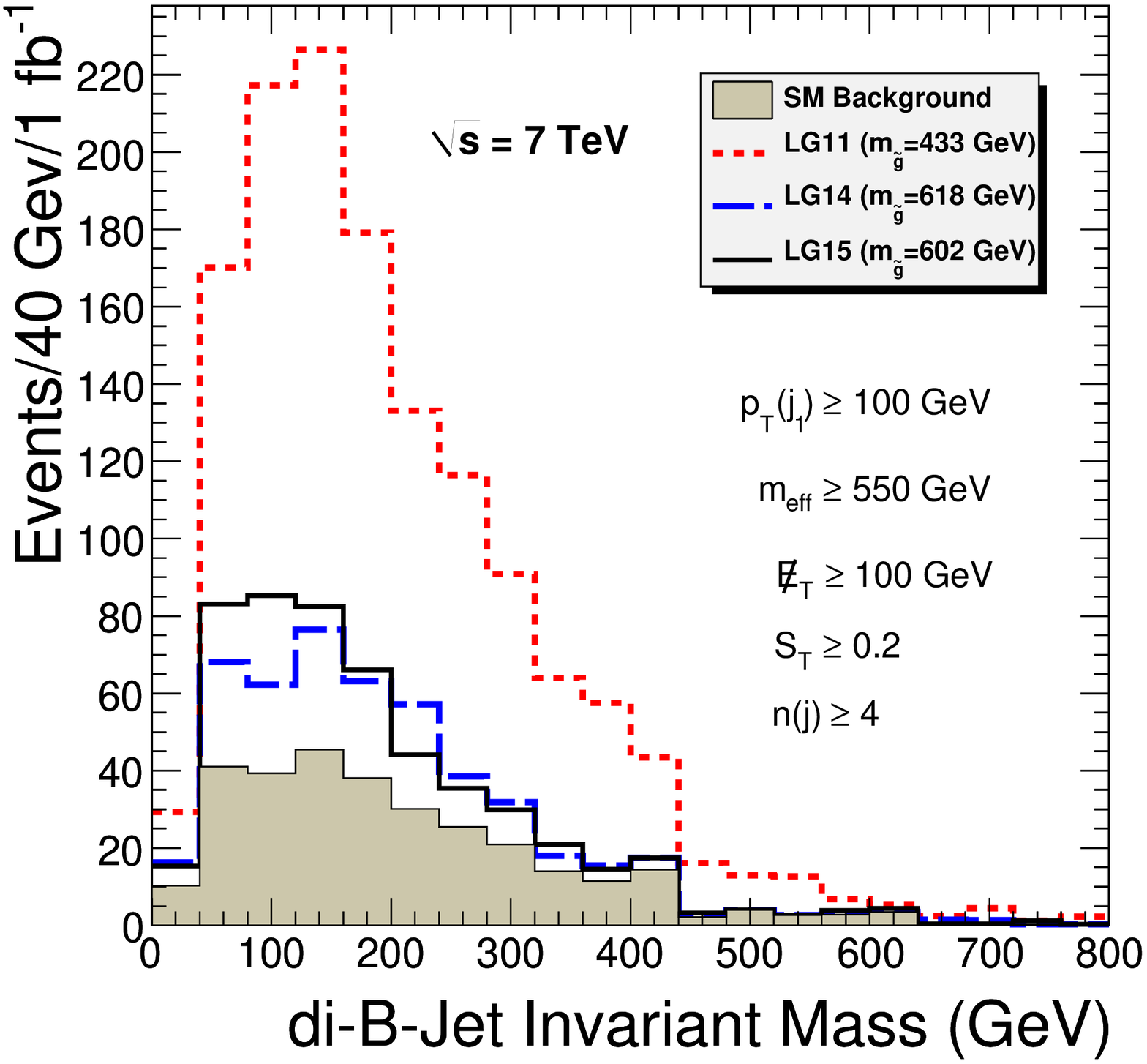}
\caption{
Left:  SUSY  plus SM background events  vs the $b$-tagged di-jet invariant mass ($m_{bb}$)
 at 1~fb$^{-1}$ of integrated luminosity 
 for signature cut $\slashed{E}_T\geq 100 \GeV$, $S_{T}\geq 0.2$, $p_{T}(j_1)\geq 100 \GeV$, $m_{\rm eff}\geq 550 \GeV$ and $n(j)\geq 4$
 for the  models LG1, LG7, LG8.
 Right: Same as the left panel except that the analysis is for 
  models LG11, LG14, LG15.   As discussed in the text, there is a hint of kinematical endpoints forming  for some of the models 
  in the di-$b$-jet invariant mass plots exhibited above.}
 \label{hist2}
\end{center}
\end{figure}

 We now discuss the reconstruction of the
$b$-tagged di-jet invariant mass peak.  In the left panel of Fig.(\ref{hist2}) we give an analysis of 
the number of SUSY event vs the $b$-tagged di-jet invariant mass ($m_{bb}$) at 1~fb$^{-1}$
 for the  models  LG1, LG7, and LG8 for the cuts displayed as well as a comparison to the SM background.  One finds that
  the three models are distinguishable above the background.  
For these models, the majority of the $b$-tagged di-jet events come from the gluino off-shell decay $\g\rightarrow\nb+b \bar b$, 
which leads to an upper bound of the kinematic endpoint $M_{bb}\le M_{\g}-M_{\nb}$ that is estimated to be in the range $(300 - 322)$~GeV.  In the left panel of Fig.(\ref{hist2}), one sees a hint of an endpoint forming in this region.  However, for these models, the kinematic endpoint is not yet discernible;
more luminosity would be needed, and further, additional uncertainties arise
in the interpretation of the invariant mass endpoint  due to additional cascade processes.  A similar analysis can be given for models LG11, LG14, and LG15 in the right panel of Fig.(\ref{hist2}). 
  The source of the jets for the three models differ from those of the left panel of Fig.(\ref{hist2}) due to their spectra. Further, 
LG11 produces a significantly larger number of  jet events compared to those for 
LG14 and LG15  due to its light color particles, i.e., the gluino and the stop, dominantly decaying to $b$~jet final states. In addition, some of the $b$~jets in the models from Fig.(\ref{hist2}) come from the light CP even
Higgs. For example, in model LG15 ${\mathcal Br} (\nb \to \na   h ) \sim 30 \%$
and  ${\mathcal Br} (h \to b \bar b ) \sim 80 \%$. Thus with increased
statistics
one may be able to partially reconstruct events coming from the
Higgs decay in this model and other models as well.

More generally in Table(\ref{tab:all})   we summarize the result of our analysis for 
  the full set of integrated luminosities 0.5~fb$^{-1}$, 1~fb$^{-1}$, 2~fb$^{-1}$, and 5~fb$^{-1}$.
     The entries in the boxes in this table 
   indicate the 
  integrated luminosity at which a  model  listed in the first column will become visible  in a specific
  signature channel listed in the top row. Thus, the entries in Table(\ref{tab:all})  show that a good  number of
  the  models in Table(\ref{bench})  will become visible at 0.5 fb$^{-1}$ of integrated luminosity, and all of the
  models given in Table(\ref{bench}) will become visible (in at least one channel)  at 5~fb$^{-1}$ of integrated luminosity.
Indeed, as discussed above,  one observes that the relative mass splitting and the relative position
  of the gluino within the sparticle mass hierarchies strongly influences the discovery capability of the LG models. Several channels
  in some cases are needed to establish a signal, and the variance amongst channels for different models is apparent.

\section{V: Conclusion\label{conclusions}}
\vspace{-.5cm}
In this work we discussed the sparticle landscape in the context of a  
 low mass gluino which is one of the prime superparticles 
that has the potential of being produced as well as
detected  at the early runs of the LHC. 
This is due to the gluino (and also squarks) being strongly interacting and  typically having 
 the largest  production cross section in $pp$ collisions at the LHC for 
sufficiently low gluino masses.  
The low mass gluino models  considered arise  in  a variety of settings
including mSUGRA, nonuniversal SUGRA models, and in supergravity/string models with a very weakly 
coupled  $U(1)^n$ extended hidden sector.
A number of specific benchmark  models were analyzed and found to be 
  encouraging for discovery at both the LHC and in dark matter experiments.
It is found that the eigencontent of the LSP in such models 
can vary over a wide range from the LSP being a pure bino, or a mixed-wino  (LG16 and LG17) 
to the LSP being  heavily Higgsino dominated (LG10). 
Further,  the associated sparticle spectrum is  found to be widely different with the squarks and
sleptons being as  low as 200 GeV (or even  less) in mass to being significantly heavy lying in the
$(2-3)$~TeV region. 
 The models analyzed  exhibit a wide array of  sparticle mass hierarchies and  
 signatures. It was shown that most of the models considered will be discoverable in the early runs
 at the LHC at $\sqrt s=7$ TeV
with 1~fb$^{-1}$  of data while all  the models will become visible with 5~fb$^{-1}$ of integrated 
luminosity. The models  considered are consistent with the  stringent bounds 
on the annihilation of neutralinos into $\gamma\gamma$ and $\gamma Z$ from 
Fermi-LAT, and further many of the models considered are discoverable in the on going dark matter experiments;
specifically 
CDMS-II,  XENON100, and EDELWEISS-2.
Further, it is found that some of  the low mass gluino  models (LG10, LG16, and LG17) can explain the 
positron excess observed in the satellite experiments such as PAMELA that  probe  antimatter in the Galaxy.
It is shown that such models  also lead to 
rich jet signatures at the LHC, thus representing a class of models 
which can be tested on multiple fronts.
\\\\
\noindent Note added :  After the publication of this paper, CMS \cite{CMSnew} and
ATLAS \cite{ATLASnew}
released their  analyses in the search for SUSY at the
LHC at $\sqrt s = 7~\rm TeV$ with 35 pb$^{-1}$.  Models LG2, and LG(11-15) have mass scales
that lie close to the observable limits quoted by ATLAS~\cite{ATLASnew}.
\\
\noindent
{\bf Acknowledgements:} 
This research is  supported in part by NSF grant
 PHY-0653342 (Stony Brook),  DOE grant  DE-FG02-95ER40899 (Michigan, MCTP),   and NSF grants  PHY-0704067 and  PHY-0757959 (Northeastern University).  Discussions and/or  communications
 with  Ben Allanach,  
  Darin Baumgartel, Gordon Kane, Ran Lu, Brent Nelson,  Jing Shao, Robert Shrock, and  Darien Wood
 are acknowledged.

\end{document}